\RequirePackage[bookmarksnumbered,unicode]{hyperref}
\documentclass[sigconf]{acmart}
\usepackage{multirow}
\usepackage{multicol}
\usepackage{amsthm}
\usepackage{algorithmic}
\usepackage{bm} 
\usepackage{booktabs}
\usepackage{enumitem}
\usepackage{amsmath,dsfont,bbm}
\usepackage{dcolumn}

\AtBeginDocument{%
  \providecommand\BibTeX{{%
    \normalfont B\kern-0.5em{\scshape i\kern-0.25em b}\kern-0.8em\TeX}}}

\acmConference[arxiv]{}{}{}

\begin{document}
\title{LLM-Driven Dual-Level Multi-Interest Modeling for Recommendation} %A Self-Reflective Large Language Model for Session-based Recommendation
\author{Ziyan Wang}
\email{wang1753@e.ntu.edu.sg}
\affiliation{%
  \institution{Nanyang Technological University}
  \country{Singapore}
}
\author{Yingpeng Du}
\email{yingpeng.du@ntu.edu.sg}
\affiliation{%
  \institution{Nanyang Technological University}
  \country{Singapore}
}
\author{Zhu Sun}
\email{zhu_sun@sutd.edu.sg}
\affiliation{%
  \institution{Singapore University of Technology and Design}
  \country{Singapore}
}
\author{Jieyi Bi}
\email{jieyi001@e.ntu.edu.sg}
\affiliation{%
  \institution{Nanyang Technological University}
  \country{Singapore}
}
\author{Haoyan Chua}
\email{haoyan001@e.ntu.edu.sg}
\affiliation{%
  \institution{Nanyang Technological University}
  \country{Singapore}
}
\author{Tianjun Wei}
\email{tianjun.wei@ntu.edu.sg}
\affiliation{%
  \institution{Nanyang Technological University}
  \country{Singapore}
}
\author{Jie Zhang}
\email{zhangj@ntu.edu.sg}
\affiliation{%
  \institution{Nanyang Technological University}
  \country{Singapore}
}

\begin{abstract}
Recently, much effort has been devoted to modeling users’ multi-interests based on their behaviors or auxiliary signals. However, existing methods often rely on heuristic assumptions,  e.g., co-occurring items indicate the same interest of users, failing to capture user multi-interests aligning with real-world scenarios. While large language models (LLMs) show significant potential for multi-interest analysis due to their extensive knowledge and powerful reasoning capabilities, two key challenges remain.
First, the granularity of LLM-driven multi-interests is agnostic, possibly leading to overly fine or coarse interest grouping. Second, individual user analysis provides limited insights due to the data sparsity issue.
In this paper, we propose an LLM-driven dual-level multi-interest modeling framework for more effective recommendation. At the user-individual level, 
we exploit LLMs to flexibly allocate items engaged by users into different semantic clusters, indicating their diverse and distinct interests. 
To alleviate the agnostic generation of LLMs, we adaptively assign these semantic clusters to users' collaborative multi-interests learned from global user-item interactions, allowing the granularity to be automatically adjusted according to the user's behaviors using an alignment module. 
To alleviate the limited insights derived from individual users' behaviors, at the user-crowd level, {we propose aggregating user cliques into synthesized users with rich behaviors} for more comprehensive LLM-driven multi-interest analysis. We formulate a max covering problem to ensure the compactness and representativeness of synthesized users' behaviors, and then conduct contrastive learning based on their LLM-driven multi-interests to disentangle item representations among different interests. 
Experiments on real-world datasets show the superiority of our approach against state-of-the-art methods.

\end{abstract}

\ccsdesc[500]{Information systems ~ Recommender systems}
\keywords{Recommender Systems, Multi-interest Modeling, Large Language Model}
\maketitle
\section{Introduction} 

\begin{figure}[t] \centering
 \includegraphics[width=0.5\textwidth]{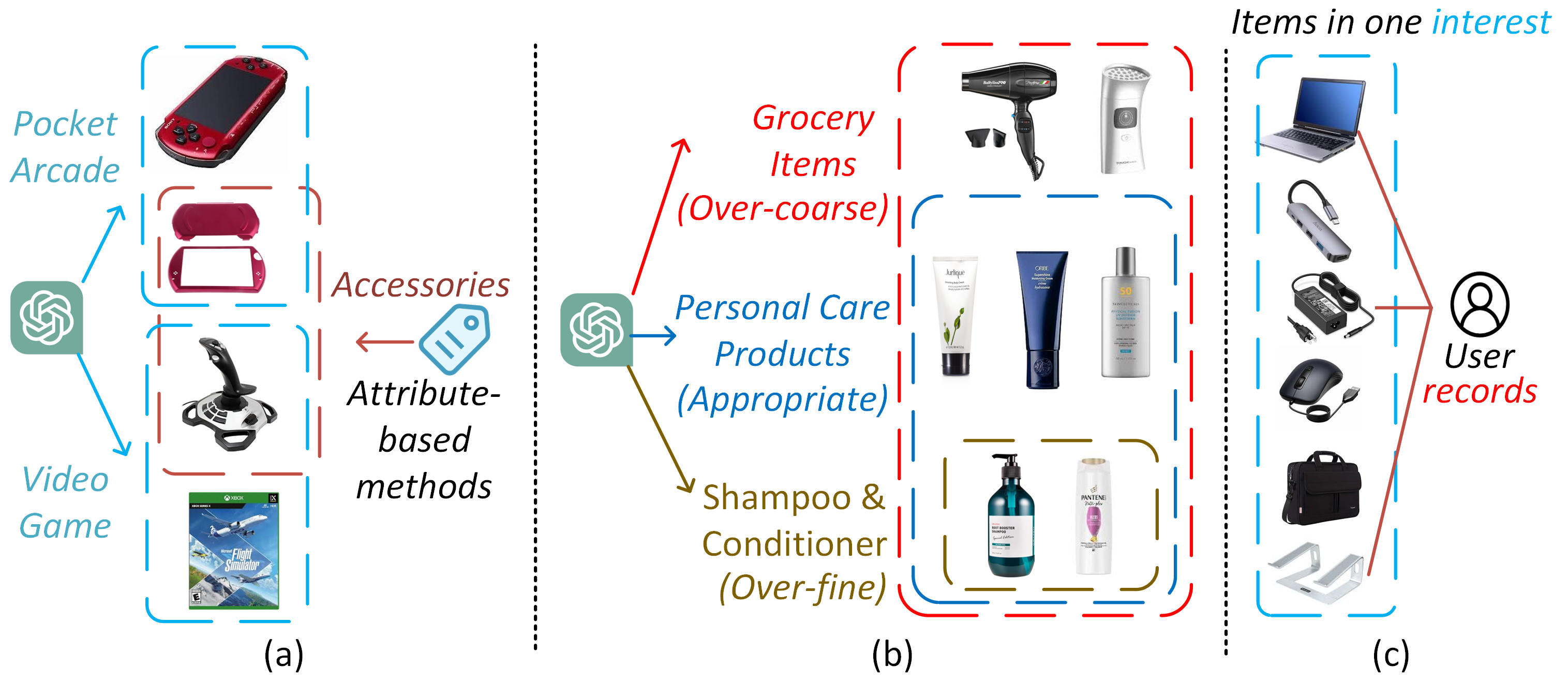}
\caption{(a) The difference between attribute-driven multi-interest analysis and LLM-driven multi-interest analysis. (b) LLM-driven multi-interest analysis across varying levels of granularity. (c) The sparsity of the user's behaviors compared to the intensive items in each interest.} \label{figintro}
\end{figure}

Recommender systems (RSs) play a crucial role in enhancing user experiences by delivering personalized content and product recommendations across various domains, including e-commerce \cite{bert4rec}, streaming services \cite{mind}, and social media \cite{social}. With the development of deep learning technologies, deep learning-based approaches have been widely adopted \cite{ncf, gru4rec,  icdemusic} to capture users' complex preferences in RSs.
Nevertheless, these methods only model the user’s preference with a single representation, overlooking the multi-faceted and discriminated nature of users' interests. To bridge this gap, multi-interest methods have gained popularity in RSs. They mainly employ capsule network \cite{mind, re4} or attention mechanism \cite{comirecsa, attenmulti} to model users' multi-interest behind their behaviors, ensuring each interest can be captured by a representation.

Despite their advancements, existing methods often rely on heuristic assumptions for multi-interest modeling, i.e., similar items indicate the same interest for users, where similarity can be measured through item embeddings \cite{ssl} or co-occurrence statistics \cite{dismir}. However, due to the inherent sparsity in user-item interactions, accurately measuring item similarity remains a significant challenge. To this end, some methods incorporate auxiliary information such as knowledge graph \cite{kgmulti}, user profile \cite{umi}, item attribute \cite{multi-attr}, and timestamps \cite{pimirec} to enhance user and item representation learning. Nonetheless, these auxiliary sources do not always guarantee accurate or comprehensive multi-interest modeling in real-world scenarios. As in Fig. \ref{figintro} (a), though the items within the red box belong to `Accessories' attribute, they significantly differ in functionalities and appeal to different user interests.
To address these problems, we leverage large language models (LLMs) with their rich knowledge and powerful reasoning capabilities, analyzing users' semantic multi-interests beyond their interaction records and auxiliary information. LLMs can offer strong explainability and semantic guidance for multi-interest extraction, providing more accurate guidance (blue boxes) on multi-interest analysis.

% Their distinctions are not identified by attribute-based methods but can be precisely captured by LLMs. 

%%if figure passed, todo: modify following examples to figure
Although leveraging LLMs offers a promising way for multi-interest modeling, directly using them is not a one-size-fits-all solution as there remain two significant challenges. 
\textbf{First}, the granularity of LLM-driven multi-interest analysis is agnostic, {\color{black} i.e., over-fine or over-coarse item division among users' behaviors, making it hard to model their' multi-interests accurately}. As in Fig. \ref{figintro} (b), LLMs might categorize interests with over-fine distinctions among items that should belong to the same interest (‘Shampoo \& Conditioner’ in the brown box), or with over-coarse groupings that fail to capture interest discrimination among items (`Grocery Items' in the red box), while only the `Personal Care Products' in the blue box is the desired or ideal interest.
\textbf{Second}, analyzing multi-interests for individual users provides limited insights due to the sparsity of users' behaviors, as relying solely on individual user interactions analyzed by LLMs may lead to incomplete modeling of multi-interests. As in Fig. \ref{figintro} (c), besides the engaged items in users' behaviors connected by red lines, there remain numerous non-co-occurring items that may also reflect the same user interests.

% it is important to capture the relationships between the related but non-co-occurring items with further utilization of LLMs.

To address the first challenge, we guide the LLM-driven multi-interest analysis at user-individual level and then adaptively align it with the learned user’s collaborative interests. Specifically, we design a prompt to guide the LLM in analyzing the user's sequentially interacted items, grouping them into distinct semantic clusters that can reflect different user interests. 
To achieve an appropriate granularity for the semantic multi-interests, we propose an alignment layer with an attentional mechanism to allocate these semantic clusters into the collaborative interests, which are learned through a capsule network based on global user-item interactions. By doing so, semantic clusters similar to a specific collaborative interest can be aggregated through attention weights, allowing the granularity to be automatically adjusted to fit the users' multi-interests, making them benefit from each other for better multi-interest modeling.

{To address the second challenge, we synthesize users with richer behaviors for LLM-driven analysis at the user-crowd level. However, synthesizing users by randomly selecting items may lead to dispersed interests of users and redundant inference of LLMs. Therefore, we provide two rules for user synthesis for LLM-driven analysis.} 
%we propose increasing the size of items for the LLM to obtain more comprehensive
On the one hand, the behaviors of synthesized users should be compact, i.e., exhibiting a limited number of interests and each interest containing rich behaviors. On the other hand, the synthesized users should be representative enough to cover as many items as possible, {contributing to a comprehensive multi-interest analysis by LLMs}. To this end, we formulate it as a max covering problem (MCP) \cite{mcp}. Specifically, to ensure the compactness, we construct synthesized users by aggregating users with similar preferences, as they are likely to share overlapping interests. To ensure the representativeness of synthesized users, we propose to maximize the coverage of their overall behaviors over items, which can be solved using a learning-based MCP solver \cite{narco}. {To make synthesized users helpful to real users' multi-interest modeling,  we conduct contrastive learning to encourage item concentration within clusters and dispersion among clusters based on LLM-driven analysis of synthesized users, eventually contributing to multi-interest learning for all users bridged by item representations in a global view.}

In summary, we propose a novel \underline{L}LM-driven \underline{D}ual-level \underline{M}ulti-\underline{I}nterest (LDMI) modeling framework to explore semantic information from both the user-individual level and user-crowd level, for more effective multi-interest recommendation.
At the user-individual level, we instruct the LLM to allocate the items engaged by each user into different semantic clusters. To alleviate the agonistic granularity, we propose to adaptively assign these semantic clusters to the collaborative interests of users. At the user-crowd level, we propose generating compact and representative synthesized users by solving an MCP, {and conducting contrastive learning to harness synthesized users for enhancing real users' multi-interest modeling}. We evaluate the proposed method through extensive experiments across three real-world datasets to demonstrate its effectiveness. Further in-depth analysis and ablation studies validate motivations and model design quantitatively and qualitatively.

\section{Literature Review}
\textbf{{Single- and Multi-interest Modeling for Recommendation}}.
%RSs have evolved significantly and can be classified into single- and multi-interest modeling for capturing user preferences.
%
Single-interest modeling in recommendation include methods built upon recurrent neural networks \cite{gru4rec}, self-attention \cite{sasrec, narm, varsa}, transformers \cite{bert4rec}, and graph convolutional networks \cite{lightgcn, icdegraph}. Some methods use data augmentation \cite{aug} or side information \cite{sideinfo} to enhance performance. However, these methods use a singular vector for user preference modeling, overlooking the distinctness of user interests and failing to capture users' multi-faceted preferences.

%\subsection{Multi-interest Modeling for Recommendation} 

Multi-interest modeling has emerged as an effective approach to improve performance. Existing methods can be classified into interaction-based and auxiliary knowledge-based based on utilized information.
The former methods solely rely on users’ engaged items. Several methods adopt capsule networks \cite{mind,remi,multimeet1} to capture users' multiple interest representations. MIND \cite{mind} first employs capsule network \cite{capsulenet} to produce multiple interest capsules through dynamic routing. Attention mechanisms are also incorporated for multi-interest modeling \cite{comirecsa, attenmulti}. Regularization strategies are adopted for multi-interest disentanglement, where Re4 \cite{re4} stabilizes the learning of multiple embeddings by incorporating backward flows, and BaM \cite{bam} promotes balanced learning across all interest vectors. DisMIR \cite{dismir} introduces an item partitioning task based on global co-purchase patterns for effective multi-task alignment. Although these methods show substantial improvements, accurately disentangling users' multi-interests remains a significant challenge due to the sparsity of user-item interactions.
The latter methods attempt to guide multi-interest learning by utilizing auxiliary knowledge such as users' profiles~\cite{umi}, timestamps \cite{pimirec}, items' categories~\cite{du2021modeling, multi-attr}, multi-type behaviors \cite{chen2021curriculum, meng2023coarse}, knowledge graphs \cite{kgmulti,wang2023intent}, and multi-modal information \cite{wang2021multimodal,wang2022disentangled}. Nonetheless, these auxiliary sources do not guarantee accurate or comprehensive multi-interest modeling in real-world scenarios.

%\subsection{Large Language Models for Recommendation}
\textbf{Large Language Models for Recommendation}.
The revolutionary success of LLMs in natural language processing has inspired researchers to incorporate LLMs into recommendation pipelines \cite{llmrecsurvey}, where related studies can be categorized as LLM-as-recommender and LLM-as-extractor.
LLM-as-recommender methods employ LLMs as scoring functions or rankers for recommendation results. Several methods fine-tune all the LLM parameters \cite{p5, transrec, elephant} or use parameter-efficient fine-tuning (PEFT) \cite{tallrec, genrec, once} to bridge the gap between LLMs and recommendation tasks. For example, \cite{tallrec} adopts LoRA \cite{lora} to enhance the LLM's generalization ability on recommendation tasks. Non-tuning methods align the recommendation objectives for LLMs without training through zero-shot prompting \cite{chatgptrs, llmrank} and in-context learning \cite{icl-llmrec, recmind}.
LLM-as-extractor methods apply LLMs for feature extraction in intermediate steps in RSs, whose outputs are subsequently fed into the recommender. Some studies focus on encoding historical behaviors and item attributes for more expressive embeddings \cite{ubert, unisrec, llmbert4rec} to capture complex semantic information within the data. Besides, several methods adopt LLMs to extract additional knowledge such as user profiles~\cite{zheng2023generative,du2023enhancing}, item descriptions~\cite{ren2024representation,wei2024llmrec}, and other textual data~\cite{liu2024once,geng2022recommendation} through semantic mining. However, these approaches overlook the distinctness of different interests of users, leaving the potential of LLMs for multi-interest modeling unexplored.

\section{Preliminaries}
%In this section, we begin with problem formulation for recommendation, followed by the background of multi-interest modeling methods and the max covering problem. 

%\subsection{Problem Formulation}
\textbf{Problem Formulation}.
In RSs, we denote the set of $M$ users as $\mathcal{U}=\{u_1,\cdots,u_M\}$ and the set of $N$ items as $\mathcal{V} = \{v_1,\cdots,v_N\}$. Each user $u_i\in\mathcal{U}$ has a sequential behavior series sorted by timestamps denoted as $s_{(u_i)} = \{v_1^i, v_2^i, \cdots, v_L^i\}$, where $v_j^i$ denotes the $j$-th item engaged by the user $u_i$ and $L$ is the length of the sequence. Besides, we suppose to know item titles in $s_{(u_i)}$ denoted as $t_{(u_i)} = \{t_1^i, t_2^i, \cdots, t_L^i\}$.  Given $s_{(u_i)}$ and $t_{(u_i)}$, we aim to infer user multi-interests and then predict items that the user is most likely to interact with at the next time step.

\textbf{Multi-Interest Modeling for Recommendation}.
Capturing different preference representations of users is essential for multi-interest methods.  Capsule networks~\cite{capsulenet} have gained popularity for multi-interest modeling recently~\cite{mind, remi}. Specifically, the capsule network $CapsuleNet(\cdot)$ can generate users’ $K$-interest representations $\{\bm{m}_1,\cdots,\bm{m}_K\}$ derived from their sequential behaviors $s_{(u_i)}$, where $K$ denotes the pre-defined number of users' multi-interests. The $k$-th collaborative interest capsule $\bm{m}_k\in \mathbb{R}^{d}$ is calculated by:
{\begin{equation}
\bm{m}_k = \sum\nolimits_{j=1}^{L} b_{jk}\cdot \bm{W}\cdot \bm{v}_j^i, \quad k = 1,\cdots,K,
\end{equation}}
where $\bm{v}_{j}^i\in \mathbb{R}^d$ denotes the embeddings of the $j$-th item in the user $u_i$'s behaviors $s_{(u_i)}$, and $d$ denotes the dimension of embedding space.  $\bm{W} \in\mathbb{R}^{d\times d}$ is the transformation matrix, and $b_{jk}$ denotes the the routing weight of item ${v}_j^i$ to the $k$-th interest capsule.
The routing weight $b_{jk}$ is calculated by the softmax operation of the routing logits $g_{jk}$ that measures the similarity between the item embedding and squashed vector $\bm{e}_k$, i.e.,
\begin{equation}
\begin{aligned}
    &\bm{e}_{k} = \frac{||\bm{m}_k||^2}{||\bm{m}_k||^2+1}\cdot\frac{\bm{m}_k}{||\bm{m}_k||}, \\
    &g_{jk} \gets g_{jk} +  {\bm{v}_j^i}^{\top}\cdot \bm{W} \cdot \bm{e}_k,\\
    &b_{jk}= \frac{exp(g_{jk})}{\sum_{k=1}^{K}exp(g_{jk})},
\end{aligned}
\end{equation}
where the iterative process updates the routing logits $g_{jk}$ and recalculates the routing weights $b_{jk}$  based on updated $\bm{m}_k$.

\textbf{Max Covering Problem}.
The max covering problem (MCP) \cite{mcp} is a combinatorial optimization problem, widely applied in decision-making under uncertainty.
Given $P$ sets, each set containing an indefinite number of objects, and $Q$ objects, each object associated with a specific value, the MCP aims to select a subset of $Z$ sets ($Z \ll P$) such that the union of $Z$ sets maximizes the sum of the associated values of the covered objects. The problem can be formulated as: 
\begin{equation}
    \begin{aligned}
        \max_{\bm{x}} &\sum\nolimits_{j=1}^Q \left(\mathbb{I} \left( \sum\nolimits_{i=1}^P \bm{x}_i \bm{A}_{ij}\ge 1 \right) \cdot w_j \right), \\
        \text{s.t.} \quad &\bm{x} \in \{0,1\}^P, \; \|\bm{x}\|_0 \leq Z,
    \end{aligned}
\end{equation}
where $\bm{A}\in\{0,1\}^{P\times Q}$ is the adjacency matrix of a bipartite graph linking the sets and objects, $w_j\in\mathbb{R}$ denotes the $j$-th object value, $\mathbb{I}(\cdot)$ is a condition indicator function, {$\bm{x}$ is the selection outcome}, and each element $\bm{x}_i$ is a scaler indicating whether the $i$-th set is selected in solution. To tackle the MCP, an advanced neural solver~\cite{narco} encodes the bipartite graph with a three-layer GraphSage model~\cite{hamilton2017inductive}, integrates the MCP constraints into a differentiable layer, and then predicts the probabilities of selecting each set.

\section{Methodology}
\begin{figure*}[t] \centering
 \includegraphics[width=0.8\textwidth]{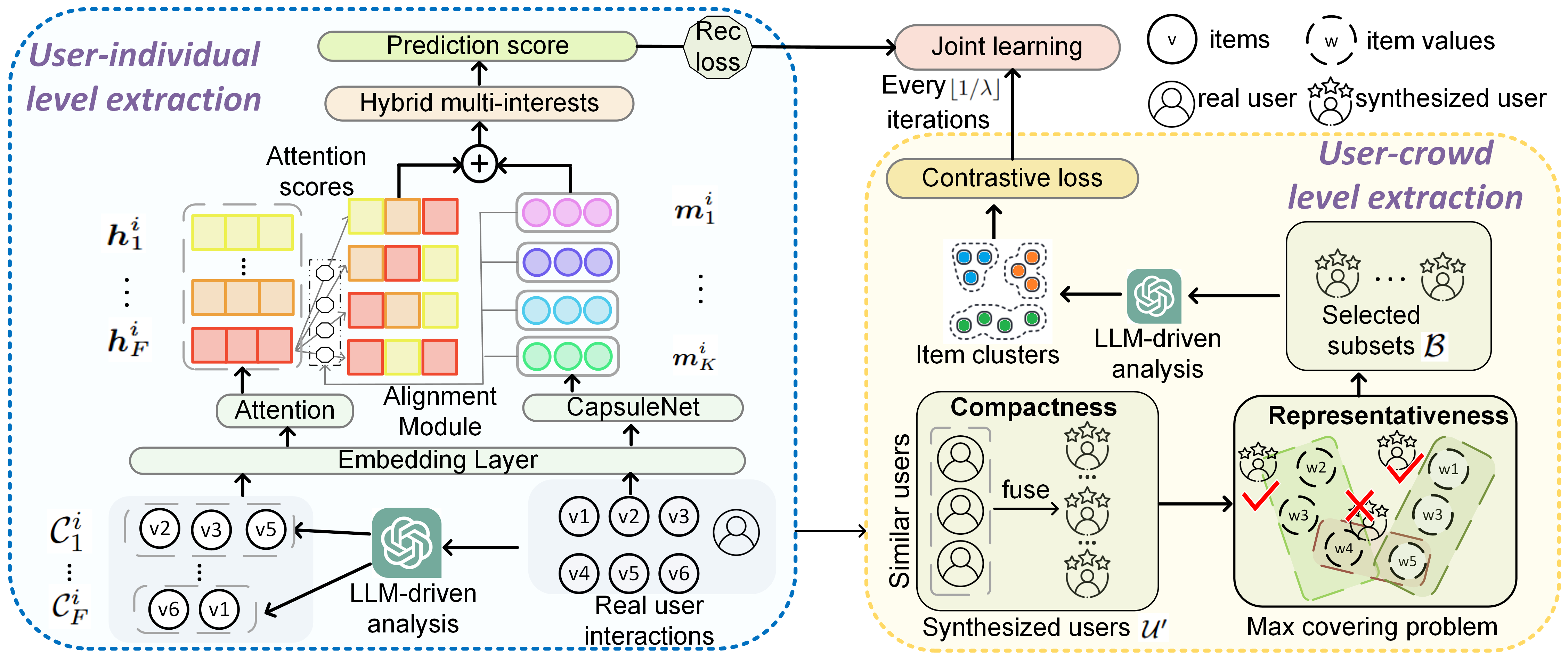}
\caption{The overall architecture of our proposed LDMI.} \label{fig:main}
\end{figure*}

This section presents \textbf{LDMI},  \underline{L}LM-based \underline{D}ual-level \underline{M}ulti-\underline{I}nterest modeling framework, which leverages LLM-driven semantics for multi-interest modeling in RSs.

\smallskip\noindent\textbf{Problem Formulation}.
In RSs, we denote the set of $M$ users as $\mathcal{U}=\{u_1,\cdots,u_M\}$ and the set of $N$ items as $\mathcal{V} = \{v_1,\cdots,v_N\}$. Each user $u_i\in\mathcal{U}$ has a sequential behavior series sorted by timestamps denoted as $s_{(u_i)} = \{v_1^i, v_2^i, \cdots, v_L^i\}$, where $v_j^i$ denotes the $j$-th item engaged by the user $u_i$ and $L$ is the length of the sequence. Besides, we suppose to know item titles in $s_{(u_i)}$ denoted as $t_{(u_i)} = \{t_1^i, t_2^i, \cdots, t_L^i\}$. Given the user's historical behavior, we aim to generate a top-n ranking list that contains items that the user is likely to engage with in the recent future.

%\subsection{Overview}
\smallskip\noindent\textbf{Model Overview}. For the user-individual level, we employ the LLM to analyze sequential behaviors for each user, infer distinctive and meaningful semantic multi-interests, and align with collaborative interests for proper granularity. For the user-crowd level, we synthesize compact and representative users with the max covering problem optimization, and bridge the gap between real and synthesized users to overcome the limited insights exhibited in individual user behaviors. Figure \ref{fig:main} shows overall framework of LDMI.

\subsection{User-individual Multi-interest Extraction} \label{localsec}
\begin{figure}[t] \centering
 \includegraphics[width=0.46\textwidth]{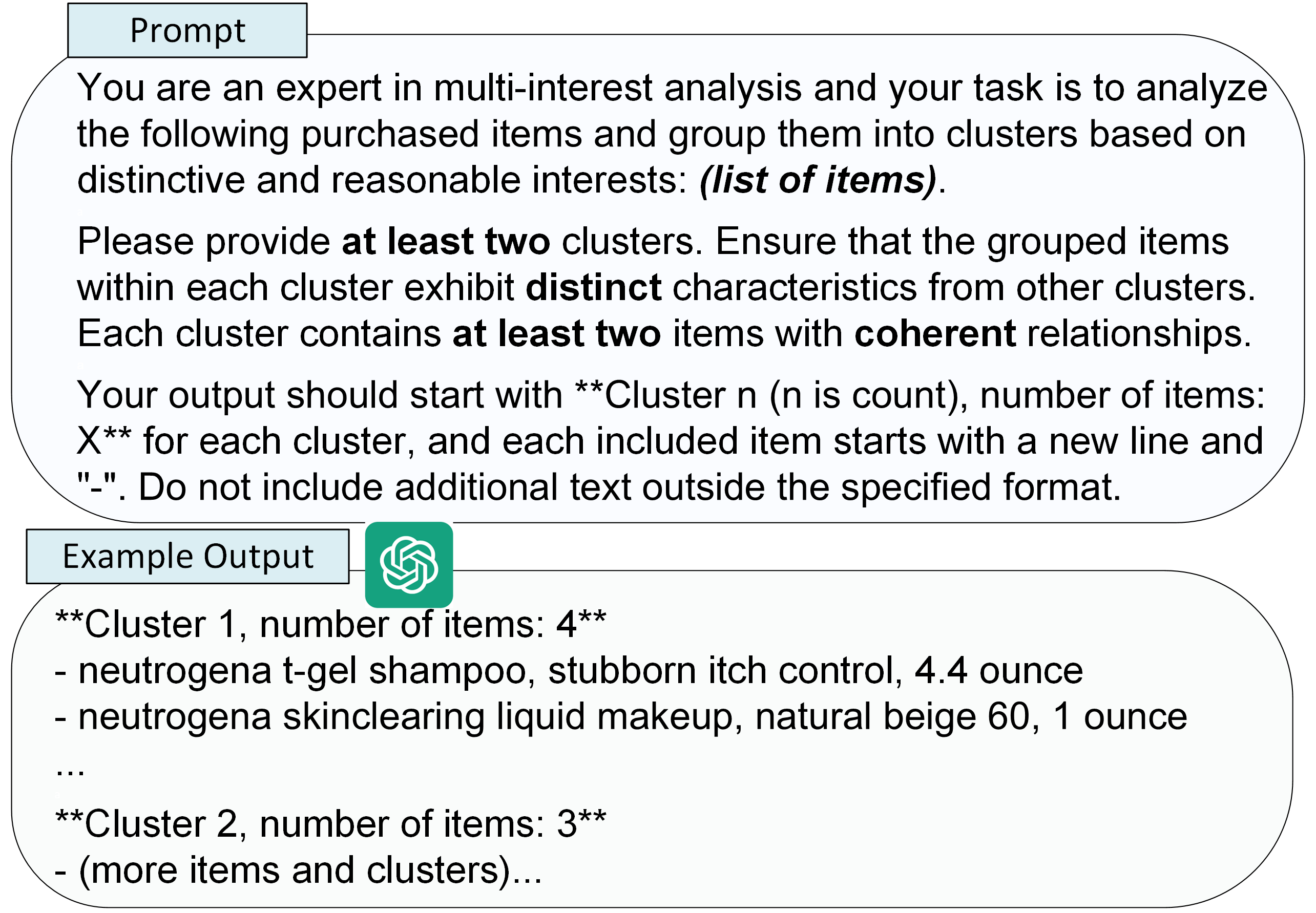}
\caption{The prompt and example output for LLM analysis.} \label{figprompt}
\end{figure}

We propose to leverage the semantic knowledge of the LLM to guide multi-interest extraction, overcoming the limitation of heuristic assumptions such as co-occurring items implying the same interest of users. 
Specifically, we prompt the LLM to conduct multi-interest analysis as follows, generating distinctive semantic interest clusters, where each representing a cohesive set of items with shared characteristics.
\begin{equation}\label{equ:llm-analysis}
    \mathcal{C}_1^i,\cdots,\mathcal{C}_{F}^i = LLM({pmt},t_{(u_i)}),
\end{equation}
where $pmt$ denotes the multi-interest analysis prompt.  $\mathcal{C}_f^i$ denotes the $f$-th cluster that contains items belonging to the same semantic group by LLM for user $u_i$. $F$ is an unknown varying number relying on LLMs' analysis and reasoning. We show the prompt and example output in Figure \ref{figprompt}. 

Intuitively, different clusters reflect distinct aspects (e.g., functionality, preference) for the user. However, the granularity of LLM-driven clusters is agnostic, making it hard to effectively model multi-interests with over-fine/coarse clusters. To address this, we propose an alignment module consisting of attention mechanism and projection layer. First,
to alleviate the over-coarse clusters in LLMs' analysis, we use an attention mechanism to dynamically aggregate the items' representations in a cluster $\mathcal{C}_f^i$, allowing more specific signals to dominate the cluster representation and sharpen the multi-interest encoding. The LLM-driven multi-interest representation is defined as:
\begin{equation} \label{eq:llmlocal}
\bm{h}_{f}^i = \sum\nolimits_{v_j \in \mathcal{C}_f^i} \alpha_j \cdot \bm{v}_j, \quad 
\alpha_j = \frac{\exp(\mathbf{w}^T \bm{v}_j + b)}{\sum_{v_k \in \mathcal{C}_f^i} \exp(\mathbf{w}^T \bm{v}_k + b)},
\end{equation}
where $\bm{v}_j$ denotes item $v_j$'s learned ID embedding, and $\mathbf{w} \in \mathbb{R}^d$, $b \in \mathbb{R}$ are learnable weights. Second, to alleviate the over-fine clusters in LLMs' analysis, we introduce an attention-based projection layer that maps semantic clusters into collaborative interests over global user-item interactions, merging clusters based on their similarity. Specifically, we adopt a capsule network to model the collaborative multi-interests:
\begin{equation}\label{eqcaps}
\bm{m}_1^i,\cdots,\bm{m}_K^i = CapsuleNet([\bm{v}_1^i,\cdots,\bm{v}_L^i]),
\end{equation}
 where $\bm{v}_j^i$ denotes the embedding of the $j$-th item engaged by the user $u_i$, and $K$ is the predefined number of users' multi-interests. Then, to align semantic clusters with collaborative interests, we compute attention scores between pairs of semantic and collaborative interest facets,

{\small\begin{equation*}
\bm{z}_k^i = \sum\nolimits_{f=1}^{F} \alpha_{kf} \cdot \bm{h}_f^i
,  \alpha_{kf} = \frac{\exp\left(\bm{m}_k^i \cdot \tanh(\bm{W_1} \bm{h}_f^i)\right)}{\sum_{f'} \exp\left(\bm{m}_k^i \cdot \tanh(\bm{W_1} \bm{h}_{f'}^i)\right)},
\end{equation*}}where $\bm{z}_k^i$ is the aggregation of the semantic clusters related to interests $\bm{m}_k^i$ of user $u_i$, $\alpha_{kf}$ is the attention score between LLM-derived {semantic clusters $\bm{h}_f^i$ and collaborative interests} $\bm{m}_k^i$, $\bm{W_1}\in\mathbb{R}^{d\times d}$ is a learnable projection matrix, and $\tanh(\cdot)$ introduces non-linearity to better capture complex cross-representation relationships.

Therefore, cluster granularity can be automatically adjusted, i.e., specific items can be emphasized through attention weights for sharper interest representation for over-coarse clusters, while similar attention weights can be learned to enable merging over-fine clusters. To take advantage of both LLM-driven semantics and collaborative information, we aggregate them into a hybrid representation: 
\begin{equation} \label{addinterest}
    \bm{o}_k^i = \bm{m}_k^i + \bm{z}_k^{i}.
\end{equation}
 Thus, we obtain the final hybrid multi-interest representations $\{\bm{o}_1^i,\cdots,\bm{o}_K^i\}$ for each user $u_i$.

\subsection{User-crowd Multi-interest Extraction} \label{globalsec}
While user-individual multi-interest modeling leverages LLMs for semantic analysis, behavioral sparsity limits its global perspective. Individual users typically interact with only a small subset of items, inevitably preventing all related items being grouped to the same interest. 
To address this, we synthesize users with richer behaviors for more comprehensive LLM analysis.
However, simply synthesizing users through random item selection produces dispersed interests and largely increases the scale and cost for LLM inference. To this end, we propose to synthesize users considering compactness and representativeness.
\textbf{Compactness} ensures synthesized users have focused interests, with each containing a cohesive set of semantically related items. Otherwise, aggregating unrelated behaviors may lead to fragmented interests and generate sparse, ineffective interest clusters.
\textbf{Representativeness} maximizes the coverage of unique items across all interests. This avoids redundant LLM analysis and enhances the generalization capability to better represent real-world interest diversity by the synthesized users.

To achieve \textbf{compactness} for user synthesis, we formulate a max covering problem (MCP) by grouping cliques of users with overlapping preferences and combine their behaviors to synthesize a user. Specifically, a clique $c_{(u_i')}$ can be generated by clustering similar users w.r.t. a real user $u_i$, i.e.,

\begin{equation}\label{eq1}
    c_{(u'_i)} = \bigcup\nolimits_{u_g \in \mathcal{N}_{(u_i)}} s_{(u_g)},
\end{equation}
where $\mathcal{N}_{(u_i)}$ denotes users who share the most overlapped behaviors to the user $u_i$. Therefore, each clique can be regarded as the union set of items of a compact synthesized user $u'_i$. We allow items to belong to multiple interest clusters, enabling LLM-driven analysis to successfully detect distinct intents. However, prompting LLM for all synthesized users leads to redundancy and inefficiency. To this end, we propose selecting representative users for LLM-driven multi-interest analysis. To further achieve \textbf{representativeness}, we select a small portion of synthesized users covering as many valuable items as possible across all interests, i.e.,
\begin{equation}\label{eq2}
\max_{|\mathcal{B}|\le Z, \mathcal{B}\subset \mathcal{U}'} \sum\nolimits_{j=1}^{N}\mathbb{I}(v_j\in\bigcup\nolimits_{u_i'\in\mathcal{B}}c_{(u_i')})\cdot w_{v_j},
\end{equation}
where $\mathcal{U}'$ is the set of synthesized users, $\mathcal{B}$ is the selected representative synthesized users with maximal size $Z$, and $w_{v_j}$ is the value for covering item $v_j$. Assuming popular items have higher values because they have more impacts, we formulate values and construct a synthesized user-item interaction matrix $\bm{A}\in\mathbb{R}^{M\times N}$ based on compactness:
\begin{equation}
w_{v_j} = 1 + \frac{\sum_{i=1}^{M}\mathbb{I}(v_j \in s_{(u_i)})}{\sum_{i=1}^{M} |s_{(u_i)}|}, \quad \bm{A}_{ij} =
\begin{cases} 
    1 &  \text{if } v_j \in c_{(u_i')}, \\
    0   & \text{otherwise},
\end{cases}
\end{equation}
where each row indicates the behaviors of a synthesized user based on the corresponding real user. We then formally formulate Equation \eqref{eq2} into a standard MCP. Specifically, we propose to represent the solution $\mathcal{B}$ with an indicator vector $\bm{x}\in\{0,1\}^M$, where $\bm{x}_i$ denotes the $i$-th element of $\bm{x}$, showing whether the synthesized user $u_i'$ is included in the solution $\mathcal{B}$. Then, Equation \eqref{eq2} can be reformulated as:
\begin{equation}
\begin{aligned}
  &\max_{|\mathcal{B}|\le Z, \mathcal{B}\subset \mathcal{U}'} \sum\nolimits_{j=1}^{N}\mathbb{I}(v_j\in\bigcup\nolimits_{u_i'\in\mathcal{B}}c_{(u_i')})\cdot w_{v_j}\\
\Leftrightarrow  &\max_{|\mathcal{B}|\le Z, \mathcal{B}\subset \mathcal{U}'} \sum\nolimits_{j=1}^{N}\mathbb{I}\left(\sum\nolimits_{u_i'\in\mathcal{B}}\mathbb{I}(v_j\in c_{(u_i')})\right)\cdot w_{v_j}\\
\Leftrightarrow  &\max_{|\mathcal{B}|\le Z, \mathcal{B}\subset \mathcal{U}'} \sum\nolimits_{j=1}^{N}\mathbb{I}\left(\sum\nolimits_{{u_i'}\in\mathcal{B}}\bm{A}_{ij}\right)\cdot w_{v_j}\\
 \Leftrightarrow & \max_{\bm{x}\in\{0,1\}^M,||\bm{x}||_0\le Z} \sum\nolimits_{j=1}^{N}\mathbb{I}\left(\sum\nolimits_{i=1}^M \bm{x}_i \cdot \bm{A}_{ij}\right)\cdot w_{v_j}.
 \end{aligned}
\end{equation}

To solve the MCP, we adopt a differentiable optimal transport model to collect the compact and representative synthesized users $\mathcal{B}$ following \cite{narco}. Given these synthesized users with rich behaviors, we propose to trigger the LLM to generate distinctive and comprehensive interest clusters for each synthetic user $u_i'\in\mathcal{B}$, i.e.,
\begin{equation}
    \mathcal{C'}_1^i,\cdots,\mathcal{C'}_{F}^i = LLM({pmt},t{(u_i')})
\end{equation}
where $t{(u_i')}$ is the titles of items within the synthesized user $u_i'$'s behaviors $c_{(u_i')}$ and $\mathcal{C'}_f^i$ is the $f$-th cluster by the LLM.

As these synthesized users do not exist in real-world RSs, directly modeling their multi-interests may have limited contributions to real users' multi-interest modeling. To this end, we model their LLM-driven multi-interests through contrastive learning to refine item distribution shared globally, {eventually contributing to real users bridged by item representations in a global view}. Specifically, we encourage item concentration within the same clusters and dispersion among different clusters in the representation space:

{\small
\begin{equation*}
\mathcal{L}^{cst}_{u_i'} = -\sum_{v_j \in c_{(u_i')}} \sum_{{v_j}*\in \mathcal{C}'(v_j)} \log \frac{ \exp\left({\bm{v}_j^\intercal\cdot {\bm{v}^*_j}}/{\tau} \right)}{\sum\limits_{v_j' \in c_{(u_i')}/\mathcal{C}'(v_j)} \exp\left({\bm{v}_j^\intercal\cdot \bm{v}_{j}'}/{\tau} \right)},
\end{equation*}}where for each synthesized user $u_i'$, $\mathcal{C}'{(v_j})$ denotes the cluster which contains the item $v_j$, $v^*_j$ denotes a positive sample that belongs to one same cluster as $v_j$, and $v_j'$ denotes negative samples that do not occur with $v_j$. $\tau$ controls the sharpness of the similarity distribution. Therefore, for each item in an interest cluster, intra-cluster items are positive instances, and inter-cluster items are hard negative instances. We aggregate the contrastive learning for all synthesized users as the overall contrastive loss,

\begin{equation}\label{gloss}
   \mathcal{L}^{cst} = - \frac{1}{|\mathcal{B}|}\sum\nolimits_{u_i' \in \mathcal{B}}\mathcal{L}^{cst}_{u_i'}.
\end{equation}

% \begin{equation} \label{eq:posneg}
% \begin{gathered}
%     \text{Pos}(q) = \bigcup_{\forall p : q \in G_{pq}} \{ j \mid j \in G_{pq} \setminus \{ q \} \}, \\
%     \text{Neg}(q) = \bigcup_{\forall p : q\in \mathcal{I}_{C_p} } \{ j \mid j \in \mathcal{I}_{C_p} \text{ and } j \notin \text{Pos}(q) \}.
% \end{gathered}
% \end{equation}

In summary, we synthesize a compact and representative subset of users with rich behaviors and formulate an MCP optimization for LLM-driven analysis, enabling global perspective for improved multi-interest modeling.

\subsection{Multi-task Objective Function}
To effectively bridge the user-individual and user-crowd multi-interest modeling, we propose a multi-task learning framework with real and synthesized users' multi-interests.
For real users, we employ hard readout following \cite{remi} to predict user-item score for recommendation:
\begin{equation}
f(u_i, v_j) = \max\nolimits_{1 \leq k \leq K} \left( {\bm{o}^i_{k}}^{\top} \bm{v}_j \right),
\end{equation}
where the interest is selected among all interests by the maximum score. For the recommendation task, the objective function can be formulated with the InfoNCE loss as:
\begin{equation} \label{recloss}
\mathcal{L}^{rec} = -\sum_{(u_i,v_j) \in \mathcal{D}} \log \frac{\exp \left( f(u_i, v_j) \right)}{\sum_{v'_j \in \mathcal{V}} \exp \left( f(u_i, v'_j)  \right)},
\end{equation}
where $\mathcal{D}$ is the training set of user-item interactions. {For selected synthesized users, we employ contrastive learning in Equation \eqref{gloss} to bridge them with real users through item representation learning to enhance multi-interest modeling.}

To conduct the multi-task learning for user-individual and user-crowd multi-interest modeling, our overall objective function aggregates these two goals in a weighted way:
\begin{equation}\label{allloss}
\mathcal{L} = \mathcal{L}^{rec} +\lambda\cdot \mathcal{L}^{cst},
\end{equation}
where $\lambda$ controls the trade-off between user-individual and user-crowd multi-interest modeling. As synthesized users usually contain rich behaviors leading to high-computation of contrastive learning, we conduct backpropagation on $\mathcal{L}^{cst}$ every $\lfloor 1/\lambda\rfloor$ iterations ($\lambda > 0$) for efficiency consideration. 

\subsection{Model Complexity and Scalability}

The computational complexity of LDMI comes from (a) LLM-driven inference and (b) multi-interest model training. For (a), assuming the complexity for analyzing each user's behaviors is $\mathcal{T}$, the total cost for user-individual multi-interest extraction is $\mathcal{O}(M\cdot\mathcal{T})$. For user-crowd multi-interest extraction, it takes $\mathcal{O}(M^2)$ for MCP and $\mathcal{O}(Z \cdot \mathcal{T})$ for LLM-driven analysis, where $Z<<M$. Therefore, the overall complexity for LLM-driven multi-interest analysis is $\mathcal{O}(M^2+ M \cdot \mathcal{T})$, which is similar to existing LLM-based recommendation methods \cite{chatgptrs, llmrank}.
Regarding (b), computing {collaborative multi-interests} takes $\mathcal{O}( L\cdot K \cdot d)$, where $L, K, d$ are sequence length, number of routing facets, embedding dimension. 
Aligning semantic and collaborative multi-interests also takes $\mathcal{O}( L\cdot K \cdot d)$. Therefore, the overall complexity of {user-individual multi-interest learning} is $\mathcal{O}(M\cdot L\cdot K \cdot d)$, which is equivalent to existing multi-interest recommendation models \cite{mind, remi}. 
For contrastive learning, it takes $\mathcal{O}(Z\cdot \overline{|c_{(u')}|}^2\cdot d)$ for each update, where $\overline{|c_{(u')}|}$ is the average number of behaviors for synthesized users.
To tackle the high computational complexity in contrastive learning, we only update the loss every $\lfloor 1/\lambda\rfloor$ iteration.
In summary, our method matches existing complexity and avoids online latency, thus ensuring scalability.

\begin{table}[t]
\centering
\setlength{\tabcolsep}{3pt}
    \centering
\caption{Dataset statistics.\label{tabledata}}
\begin{tabular}{c| c c c c c}
\toprule
% Column headers for baseline methods
\textbf{Dataset}& \# Users & \# Items & \# Interactions & Avg Len &Density  \\
\midrule
Beauty &15,097&44,261&100,055&6.63&1.5e-4\\
Book &99,101&361,002&780,018&7.87&2.2e-5\\
Game &20,551&27,456&153,541&7.47&2.7e-4\\
\bottomrule
\end{tabular}
\end{table}

\begin{table*}[t]
    \centering
    \caption{Performance comparison of baseline methods and our proposed LDMI on three datasets. The best results are in \textbf{bold} and the runner-up results are \underline{underlined}. The improvements are significant on the t-test ($p \leq 0.01$) for all the
metrics.}
\begin{tabular}{l |c| c c c  c  c c c  c c c c}
\toprule
% Column headers for baseline methods
& Metrics & Pop & GRU4Rec   & LLMBRec & MIND & ComiRec & Re4 & REMI & DisMIR & EIMF & \textbf{LDMI}  \\
\midrule

% Group: Beauty
\multirow{6}{*}{\rotatebox{90}{\textbf{Beauty}}} 
& \textit{R@20} & 0.0228 & 0.0349 & 0.0289& 0.0477 & 0.0367 & 0.0550 & 0.0616 & 0.0702 & \underline{0.0765}& \textbf{0.0872} \\
& \textit{N@20} & 0.0161 & 0.0180  &0.0205& 0.0248 & 0.0176 & 0.0271 & 0.0320 & 0.0364 & \underline{0.0390}& \textbf{0.0443}\\
& \textit{H@20} & 0.0351 & 0.0454  &0.0580& 0.0669 & 0.0500 & 0.0715 & 0.0838 & 0.1057& \underline{0.1126}& \textbf{0.1380}  \\
& \textit{R@50} & 0.0391 & 0.0452  &0.0473& 0.0646 & 0.0519 & 0.0751 & 0.0817 & 0.0955& \underline{0.0982}& \textbf{0.1092} \\
& \textit{N@50} & 0.0209 & 0.0186 &0.0272& 0.0257 & 0.0194 & 0.0274 & 0.0325 & \underline{0.0433}& 0.0427& \textbf{0.0505}  \\
& \textit{H@50} & 0.0593 & 0.0613 &0.0943& 0.0897 & 0.0702 & 0.0987 & 0.1099 &0.1360& \underline{0.1429}& \textbf{0.1552}\\
\midrule

% Group: Book
\multirow{6}{*}{\rotatebox{90}{\textbf{Book}}} 
& \textit{R@20} & 0.0075 & 0.0215 &0.0187&0.0236 & 0.0275 & 0.0298 & 0.0441 & 0.0639 & \underline{0.0722}& \textbf{0.0804}  \\
& \textit{N@20} & 0.0052 & 0.0112  &0.0132& 0.0154 & 0.0166 & 0.0187 & 0.0293 & 0.0401 & \underline{0.0413}&\textbf{0.0485} \\
& \textit{H@20} & 0.0121 &0.0314  &0.0311&0.0334 & 0.0382 & 0.0420 &0.0639  &0.1034& \underline{0.1060}& \textbf{0.1228}  \\
& \textit{R@50} &0.0133 & 0.0296 &0.0256&0.0313  &0.0381  & 0.0425& 0.0592&0.0898& \underline{0.0951}& \textbf{0.1036} \\
& \textit{N@50} & 0.0070 & 0.0113 &0.0148& 0.0158& 0.0169 & 0.0194 &0.0305 &0.0404& \underline{0.0443}&\textbf{0.0519} \\
& \textit{H@50} &  0.0217 & 0.0431  &0.0538& 0.0482 & 0.0570 & 0.0630 & 0.0915 & 0.1367& \underline{0.1436} &\textbf{0.1594} \\
\midrule

% Group: Game
\multirow{6}{*}{\rotatebox{90}{\textbf{Game}}} 
& \textit{R@20} & 0.0226 & 0.0751 &0.0661&0.0950 &0.0751 & 0.0967 & 0.1082 & \underline{0.1221}& 0.1172 & \textbf{0.1305}  \\
& \textit{N@20} & 0.0139 & 0.0393  &0.0420&0.0514&0.0384  &0.0533  &0.0543  &\underline{0.0618}  & 0.0604 &\textbf{0.0713}   \\
& \textit{H@20} &  0.0372 & 0.1099     & 0.1137& 0.1401 & 0.1050  & 0.1465  &0.1571   &\underline{0.1689} & 0.1593 & \textbf{0.2133} \\
& \textit{R@50} &   0.0435& 0.1073    &0.0849&0.1387  &0.1145   &0.1409   &  0.1510 &\underline{0.1597}  & 0.1571&\textbf{0.1742} \\
& \textit{N@50} & 0.0206  &0.0423     &0.0445& 0.0552 & 0.0401  & 0.0558  & 0.0581  &\underline{0.0689}& 0.0627 &\textbf{0.0774}   \\
& \textit{H@50} &  0.0710 & 0.1552    &0.1576&0.2048   &0.1496   &0.2007   & 0.2175  & \underline{0.2479}& 0.2315& \textbf{0.2662}  \\
\bottomrule
\end{tabular}
\label{tablemainres}
\end{table*}
\section{Experiments}

In this section, we validate the effectiveness of LDMI through experiments and answer the following research questions.
\textbf{RQ1}: Can LDMI outperform state-of-the-art multi-interest recommendation methods?
\textbf{RQ2}: Whether LDMI benefits from user-individual and user-crowd multi-interest modeling with LLM-driven analysis?
\textbf{RQ3}: How do hyper-parameters and {MCP solver selection} affect the performance of LDMI?
\textbf{RQ4}: How does LDMI outperform baselines in multi-interest extraction qualitatively?

\subsection{Experimental Setup}
\subsubsection{Datasets} We use three subcategories of Amazon Review Data \cite{amazonreview} with varying scales, namely \textit{Beauty}, \textit{Books}, and \textit{Video Games}, abbreviated as Beauty, Book, and Game, respectively. All three datasets contain users' ratings on items with timestamps and item title information. Following prior studies \cite{remi, dismir}, we filter out users and items with less than 5 records, then we convert ratings as implicit feedback for these three datasets. The details are summarized in Table \ref{tabledata}.

\subsubsection{Evaluation Settings}
To ensure a fair comparison, we follow prior studies \cite{remi}, we chronologically split the user interactions with maximum length of 20 into training, validation, and test sets by the proportion of 6:2:2 and test last 20\% items in each sequence. We adopt three widely used top-$n$ evaluation metrics, i.e., Recall (R), Hit Rate (H), and Normalized Discounted Cumulative Gain (N), to evaluate all methods with $n = \{20,50\}$. Experimental results are reported as the average of five runs for the significance test.

\subsubsection{Baseline Methods}
We compare our model LDMI with the following baselines.
    \textbf{Pop} takes the most popular items as the recommendation results.
    \textbf{GRU4Rec} \cite{gru4rec} models sequential behaviors through RNN structure.
    \textbf{LB4Rec} \cite{llmbert4rec} leverages an LLM to produce expressive embeddings by the BERT4Rec structure. 
     \textbf{MIND} \cite{mind} uses dynamic routing with a capsule network for multi-interest learning.
     \textbf{ComiRec-SA} \cite{comirecsa} allows diversity control and introduces multi-head attention to model users' multi-interests.
     \textbf{Re4} \cite{re4} leverages the backward flow to re-examine and regulate interest representations.
     \textbf{REMI} \cite{remi} introduces interest-aware hard negative sampling with routing variation regularization for multi-interest learning.
     \textbf{DisMIR} \cite{dismir} formulates an item partition problem to encourage items in each group to focus on a discriminated interest.
     \textbf{EIMF} \cite{qiao2024llm} uses an LLM to extract similar items for multi-interest modeling.

\subsubsection{Implementation Details}
For a fair comparison, all methods are optimized by the Adam optimizer with a batch size of 128 and we adopt fixed embedding dimension 64 for all methods following \cite{remi, dismir}. For all methods, we select the best performance by varying the number of interests in \{$2, 4, 6, 8$\}, learning rate in \{$1e^{-2}, 1e^{-3}, 1e^{-4}$\}, and weight decay in \{$1e^{-4}, 1e^{-5}, 1e^{-6}$\}. For hyper-parameters, we set the temperature $\tau=0.1$ and $\lambda=0.01$ for the contrastive loss $\mathcal{L}^{cst}$. For the MCP solver, we follow the same hyper-parameters as suggested in the original paper \cite{narco}. For the LLM implementation, we use the \textit{gpt-4o} as the backbone model with temperature set to 0 for reproducibility. For other hyper-parameters in baseline methods, we follow the authors' implementation if they exist, otherwise we tune them to their best according to their performance on the validation set. 
All methods are run on one NVIDIA GeForce RTX 2080 Ti GPU with 11GB memory and 4 GHz CPU.

\begin{table}[t]
\caption{Experimental results of ablation studies.}
\centering
\begin{tabular}{ c | c c | c c | c }
\toprule
  Metrics & w/o-sem & w/o-col & w/o-com & w/o-rep  & LDMI \\
\midrule

 \textit{R@20} & 0.0544 &0.0604&0.0841 &0.0850&\textbf{0.0872}\\
 \textit{N@20} & 0.0272 &0.0310&0.0403 &0.0425&\textbf{0.0443}\\
 \textit{H@20} & 0.0745 &0.0920&0.1310 &0.1344&\textbf{0.1380}\\
 \textit{R@50} & 0.0747 &0.0889&0.1042 &0.1058&\textbf{0.1092}\\
 \textit{N@50} & 0.0285 &0.0338&0.0471 &0.0479&\textbf{0.0505}\\
\textit{H@50} & 0.1013 &0.1279&0.1473 &0.1525&\textbf{0.1552}\\
\bottomrule
\end{tabular}
\label{tableablationa}
\end{table}

\subsection{Results and Analysis}
%\subsection{Comparison with Baselines (RQ1)}
\subsubsection{Comparison with Baselines (RQ1)}
From the results in Table \ref{tablemainres}, we summarize our key findings to answer RQ1. 
{First}, LDMI consistently outperforms baselines across all three datasets, highlighting the effectiveness of our LLM-driven analysis. In addition, the improvements demonstrate the advantage of integrating LLM analysis with our dual-level multi-interest modeling framework by capturing diverse and meaningful user interests. We mitigate issues of agnostic granularity in LLM generation and limited insights based on sparse behaviors of individual users.
{Second}, we observe that multi-interest baselines generally outperform single-interest baselines, showing that capturing multiple facets of user interests is beneficial for better results.
{Third}, the superior performance of LDMI over the LLM-based baseline demonstrates that our approach achieves effective and coherent integration of the LLM-driven analysis and conventional multi-interest recommendation methods. Specifically, LDMI outperforms the relatively strong EIMF due to its emphasize on personalization and the granularity issue.
{Last}, the relatively strong performance of DisMIR shows the positive impact of a global perspective item partition task. However, relying solely on the sparse co-occurrence of items limits the insights about item relationships. In comparison, LDMI produces synthesized users with compact and representative item subsets by formulating and solving the MCP.

\begin{figure*}[t]
    \centering
    \begin{tabular}{ccc} % 4 columns
        \begin{minipage}[b]{0.27\textwidth}
            \centering
            \includegraphics[width=\textwidth]{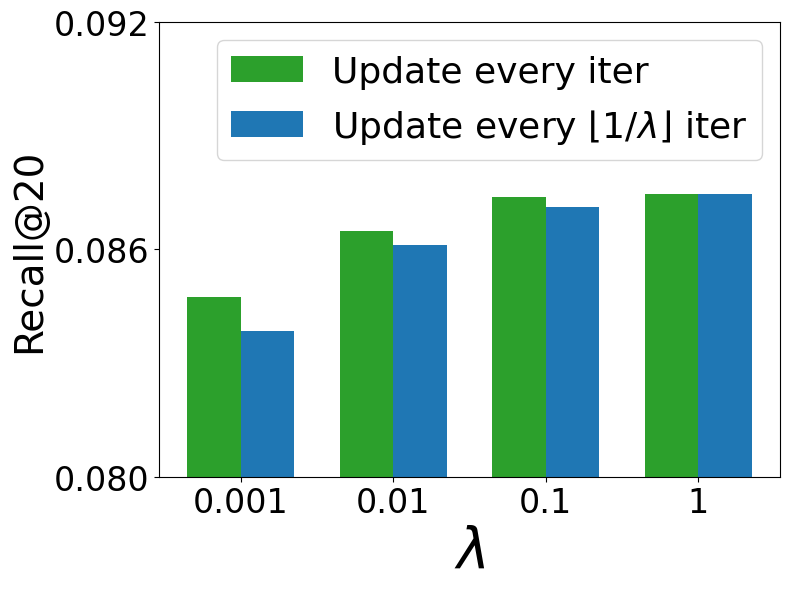}
        \end{minipage} &
        \begin{minipage}[b]{0.27\textwidth}
            \centering
            \includegraphics[width=\textwidth]{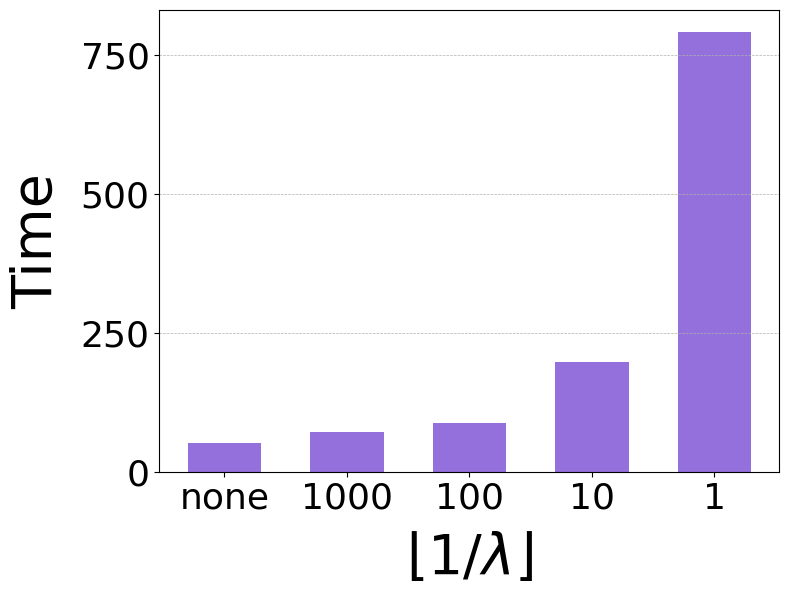}
        \end{minipage} &
        \begin{minipage}[b]{0.27\textwidth}
            \centering
            \includegraphics[width=\textwidth]{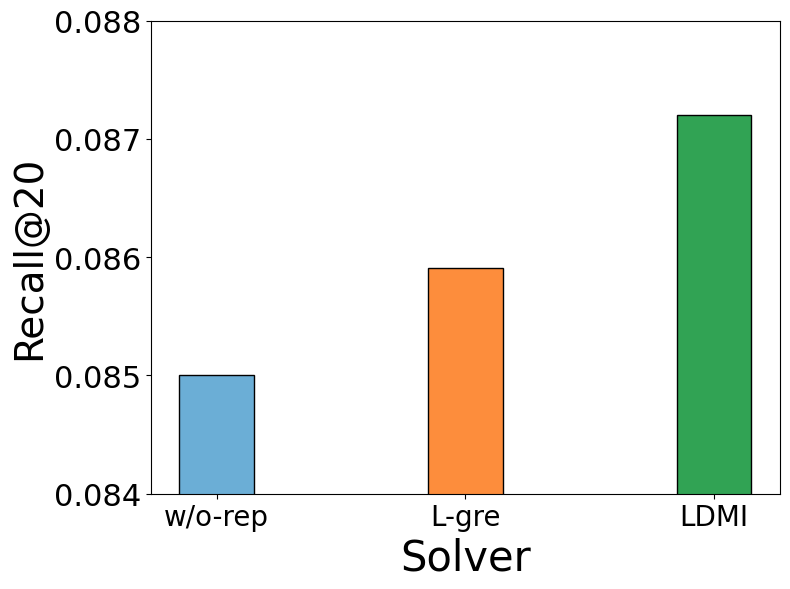}
        \end{minipage}\\
    \end{tabular}
    \caption{The model performance (a) and training time (b) with varying user-crowd level loss weights and MCP solver (c).}
    \label{interval}
\end{figure*}

\subsubsection{Ablation Studies (RQ2)}
To validate the effects of key components, we conduct ablation studies as follows. \textbf{w/o-sem} removes the \underline{sem}antic-based multi-interest modeling with LLM-driven analysis at user-individual level, i.e., $\bm{h}_f^i=0$.
\textbf{w/o-col} removes the \underline{col}laborative-based multi-interest modeling at the user-individual level, i.e., $\bm{o}_k^i=\bm{h}_k^i$.
\textbf{w/o-com} removes the \underline{com}pactness rule for user synthesis, e.g., MCP formulation and user synthesize based on users' similarities. 
 \textbf{w/o-rep} removes the \underline{rep}resentativeness rule for user synthesis. Instead, it randomly selects cliques as synthesized users for LLM-driven analysis at the user-crowd level.

 From Table \ref{tableablationa}, first, at the user-individual level, the lack of semantic multi-interests (w/o-sem) results in the worst performance, proving the effectiveness of LLM-driven analysis in handling the limitations of existing multi-interest modeling assumptions like co-occurring items simply indicating same interests.
 Second, w/o-col also shows inferior performance, showing our integration with collaborative interests and the alignment module is successful in alleviating the agnostic granularity issue in LLM-driven multi-interest analysis. 
 Third, w/o-com and w/o-rep shows degraded performance, indicating that analyzing multi-interests for individual users only provides limited insights due to the sparsity of user behaviors. On the one hand, generating user cliques from similar preferences ensures a moderate number of interests for synthesized users, thus each containing a rich set of cohesive items. On the other hand, the formulated MCP can reduce redundancy and enhance representativeness. Thus, LDMI achieves more comprehensive multi-interest modeling through our crowd-level multi-interest extraction with synthesized users and the MCP optimization.

\begin{table}[t]
\centering
\fontsize{9}{9}\selectfont
    \centering
    \caption{LDMI performance of different interest numbers.}
\begin{tabular}{l |c| c c c c}
\toprule
% Column headers for baseline methods
& Metrics & 2 & 4 & 6 & 8  \\
\midrule

% Group: Beauty
\multirow{6}{*}{\rotatebox{90}{\textbf{Beauty}}} 
& \textit{R@20} &0.0869&0.0872&0.0847&0.0849\\
& \textit{N@20} &0.0438&0.0443&0.0423& 0.0417\\
& \textit{H@20} &0.1374&0.1380&0.1295& 0.1334\\
& \textit{R@50} &0.1077&0.1092&0.1009&0.1006\\
& \textit{N@50} &0.0497&0.0505&0.0469&0.0486 \\
& \textit{H@50} & 0.1530 & 0.1552 & 0.1523 & 0.1537\\
\bottomrule
\end{tabular}
\label{tableinterestnum}
\end{table}

\subsubsection{Hyper-parameter Analysis (RQ3)}
 Table \ref{tableinterestnum} shows the effect of the number of multi-interests $K$ for LDMI. Specifically, insufficient interest numbers make it hard to capture the diverse facets of user preferences, while too large interest numbers may lead to over-fine multi-interest modeling. We suggest moderate interest number 4.

{Fig. \ref{interval}(a) investigates the impact of the loss weight $\lambda$ on model accuracy and training time (b), with an update conducted every $\lfloor 1/\lambda\rfloor$ iterations. Higher $\lambda$ (lower $\lfloor 1/\lambda\rfloor$) usually leads to higher model accuracy but requires a significantly longer time for model training. To balance effectiveness and efficiency, we suggest selecting $\lfloor 1/\lambda\rfloor=100$.} Generally, we observe that their performance improves as $\lambda$ increases, with only slight differences between the two update strategies when $\lambda \geq 1 \times 10^{-2}$. Therefore, the efficient update strategy with a proper setting of $\lambda$ significantly accelerates the training while maintaining comparable performance.

In Fig. \ref{interval}(c), we test the performance of different solvers in MCP for the multi-interest extraction at the user-crowd level. Specifically, we replace the employed MCP solver \cite{narco} with a greedy search of representative synthesized users, L-gre.  First, LDMI shows an advantage over L-gre, reflecting the effectiveness of the MCP solver in selecting representative synthesized users. Second, both of the two approaches (LDMI and L-gre) outperform the w/o-rep variant, indicating the necessity of {selecting representative synthetic users} at the user-crowd level. 

\begin{figure}[t] \centering
 \includegraphics[width=0.46\textwidth]{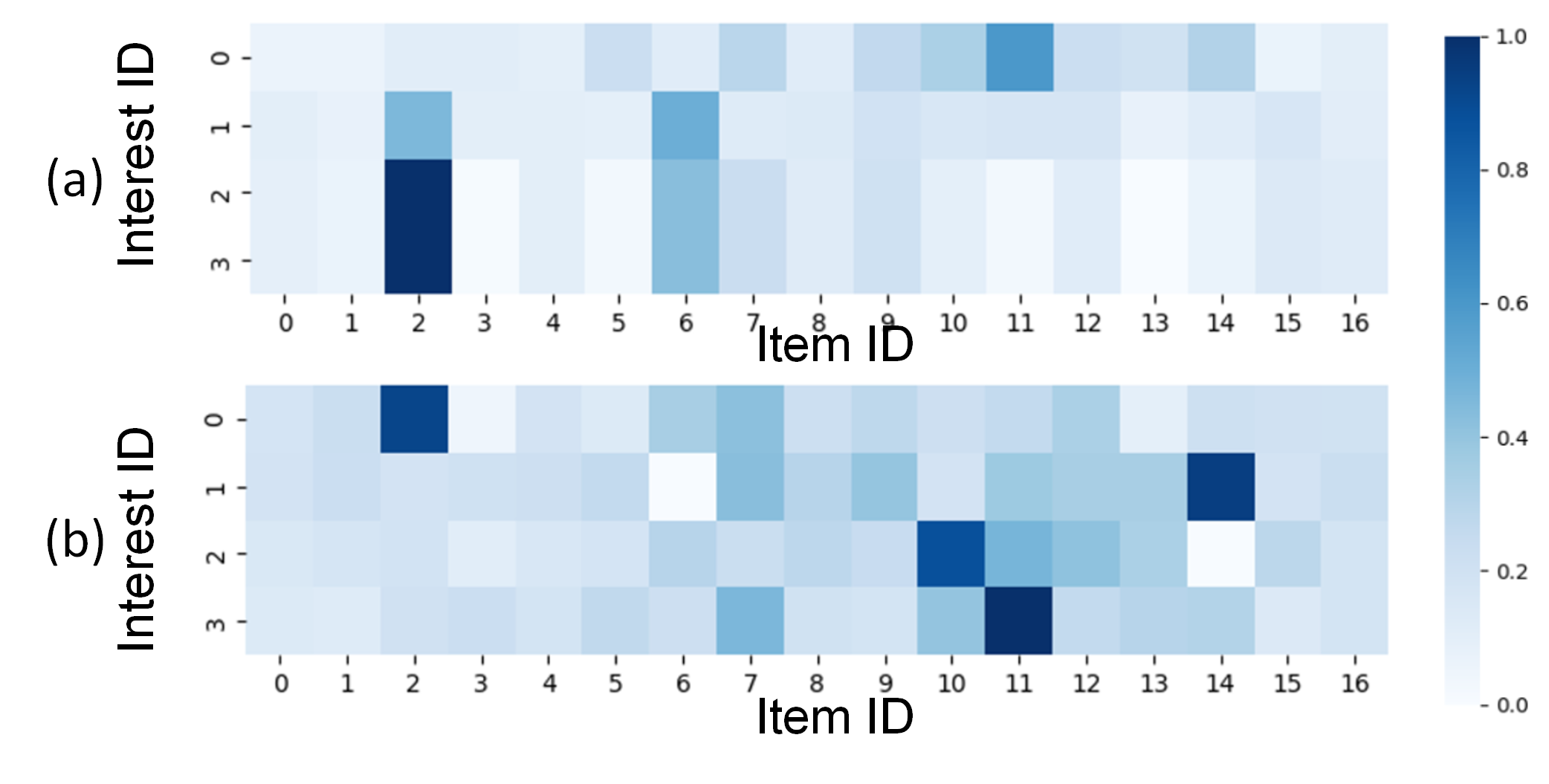}
\caption{Multi-interest heatmap of
MIND (a), LDMI (b).
} \label{fig:case-local}
\end{figure}

\subsubsection{Case Study (RQ4)} \label{sec:case}
We illustrate a case study to qualitatively investigate the effect of our model for multi-interest extraction via LLMs. To visualize the cosine similarity between interests and items, we plot the multi-interest heatmap in Fig. \ref{fig:case-local} for a classic baseline MIND and our LDMI model, which corresponds to a user sequence randomly sampled from \textit{Beauty} dataset containing 17 items across 4 interests. In the heatmap, darker colors indicate higher cosine similarity between interests and items.
 We notice that the heatmaps with interest ID 2 and 3 in MIND exhibit significant overlap, indicating collapsed facets. Rather than being dominated mainly by one item for each interest in MIND, each interest in LMDI contains multiple relevant items, indicating discriminated and balanced multi-interest modeling. This improvement validates LDMI’s ability to capture diverse and fine-grained interests with the help of the LLM, contributing to better performance for recommendation.

\section{Conclusion}
In this paper, we present LDMI, the LLM-based Dual-level Multi-Interest approach that leverages the semantic knowledge and reasoning skills of LLMs to address critical challenges in multi-interest {modeling for} recommendation. 
At the user-individual level, LDMI resolves the agnostic granularity in LLM-driven analysis by aligning semantic clusters with collaborative interests via an alignment module, achieving effective multi-interest representation extraction.
At the user-crowd level, LDMI aggregates user cliques into synthesized users with rich behaviors for LLM analysis to alleviate the sparsity of individual user behaviors. We leverage contrastive learning to utilize synthesized users for enhancing real users’ multi-interest modeling.
Extensive experiments validate the superiority of LDMI over single- and multi-interest baselines as well as LLM-based methods. Further analysis supports the effectiveness of {our dual-level design and LDMI's ability to address key limitations of prior approaches.}
{In future work, we will extend LDMI by fine-tuning open-source LLMs to align semantics with collaborative behaviors without the reliance on external APIs, improving scalability for multi-interest modeling}.

\bibliographystyle{ACM-Reference-Format}
\bibliography{output.bib}


\begin{thebibliography}{58}

%%% the Web address.

\ifx \showCODEN    \undefined \def \showCODEN     #1{\unskip}     \fi
\ifx \showDOI      \undefined \def \showDOI       #1{#1}\fi
\ifx \showISBNx    \undefined \def \showISBNx     #1{\unskip}     \fi
\ifx \showISBNxiii \undefined \def \showISBNxiii  #1{\unskip}     \fi
\ifx \showISSN     \undefined \def \showISSN      #1{\unskip}     \fi
\ifx \showLCCN     \undefined \def \showLCCN      #1{\unskip}     \fi
\ifx \shownote     \undefined \def \shownote      #1{#1}          \fi
\ifx \showarticletitle \undefined \def \showarticletitle #1{#1}   \fi
\ifx \showURL      \undefined \def \showURL       {\relax}        \fi
% invisible to TeX
\providecommand\bibfield[2]{#2}
\providecommand\bibinfo[2]{#2}
\providecommand\natexlab[1]{#1}
\providecommand\showeprint[2][]{arXiv:#2}

\bibitem[Bai et~al\mbox{.}(2018)]%
        {sideinfo}
\bibfield{author}{\bibinfo{person}{Ting Bai}, \bibinfo{person}{Jian-Yun Nie}, \bibinfo{person}{Wayne~Xin Zhao}, \bibinfo{person}{Yutao Zhu}, \bibinfo{person}{Pan Du}, {and} \bibinfo{person}{Ji-Rong Wen}.} \bibinfo{year}{2018}\natexlab{}.
\newblock \showarticletitle{An Attribute-aware Neural Attentive Model for Next Basket Recommendation}. In \bibinfo{booktitle}{\emph{The 41st International ACM SIGIR Conference on Research \& Development in Information Retrieval (SIGIR)}}. \bibinfo{pages}{1201–1204}.
\newblock


\bibitem[Bao et~al\mbox{.}(2023)]%
        {tallrec}
\bibfield{author}{\bibinfo{person}{Keqin Bao}, \bibinfo{person}{Jizhi Zhang}, \bibinfo{person}{Yang Zhang}, \bibinfo{person}{Wenjie Wang}, \bibinfo{person}{Fuli Feng}, {and} \bibinfo{person}{Xiangnan He}.} \bibinfo{year}{2023}\natexlab{}.
\newblock \showarticletitle{TALLRec:An Effective and Efficient Tuning Framework to Align Large Language Model with Recommendation}. In \bibinfo{booktitle}{\emph{Proceedings of the 17th ACM Conference on Recommender Systems (RecSys)}}. \bibinfo{pages}{1007–1014}.
\newblock


\bibitem[Cen et~al\mbox{.}(2020)]%
        {comirecsa}
\bibfield{author}{\bibinfo{person}{Yukuo Cen}, \bibinfo{person}{Jianwei Zhang}, \bibinfo{person}{Xu Zou}, \bibinfo{person}{Chang Zhou}, \bibinfo{person}{Hongxia Yang}, {and} \bibinfo{person}{Jie Tang}.} \bibinfo{year}{2020}\natexlab{}.
\newblock \showarticletitle{Controllable Multi-Interest Framework for Recommendation}. In \bibinfo{booktitle}{\emph{Proceedings of the 26th ACM SIGKDD International Conference on Knowledge Discovery \& Data Mining (KDD)}}. \bibinfo{pages}{2942–2951}.
\newblock


\bibitem[Chai et~al\mbox{.}(2022)]%
        {umi}
\bibfield{author}{\bibinfo{person}{Zheng Chai}, \bibinfo{person}{Zhihong Chen}, \bibinfo{person}{Chenliang Li}, \bibinfo{person}{Rong Xiao}, \bibinfo{person}{Houyi Li}, \bibinfo{person}{Jiawei Wu}, \bibinfo{person}{Jingxu Chen}, {and} \bibinfo{person}{Haihong Tang}.} \bibinfo{year}{2022}\natexlab{}.
\newblock \showarticletitle{User-Aware Multi-Interest Learning for Candidate Matching in Recommenders}. In \bibinfo{booktitle}{\emph{Proceedings of the 45th International ACM SIGIR Conference on Research and Development in Information Retrieval (SIGIR)}}. \bibinfo{pages}{1326–1335}.
\newblock


\bibitem[Chen et~al\mbox{.}(2021b)]%
        {pimirec}
\bibfield{author}{\bibinfo{person}{Gaode Chen}, \bibinfo{person}{Xinghua Zhang}, \bibinfo{person}{Yanyan Zhao}, \bibinfo{person}{Cong Xue}, {and} \bibinfo{person}{Ji Xiang}.} \bibinfo{year}{2021}\natexlab{b}.
\newblock \showarticletitle{Exploring Periodicity and Interactivity in Multi-Interest Framework for Sequential Recommendation.}. In \bibinfo{booktitle}{\emph{Proceedings of the 30th International Joint Conference on Artificial Intelligence (IJCAI)}}. \bibinfo{pages}{1426--1433}.
\newblock


\bibitem[Chen et~al\mbox{.}(2021a)]%
        {chen2021curriculum}
\bibfield{author}{\bibinfo{person}{Hong Chen}, \bibinfo{person}{Yudong Chen}, \bibinfo{person}{Xin Wang}, \bibinfo{person}{Ruobing Xie}, \bibinfo{person}{Rui Wang}, \bibinfo{person}{Feng Xia}, {and} \bibinfo{person}{Wenwu Zhu}.} \bibinfo{year}{2021}\natexlab{a}.
\newblock \showarticletitle{Curriculum disentangled recommendation with noisy multi-feedback}.
\newblock \bibinfo{journal}{\emph{Advances in Neural Information Processing Systems (NeurIPS)}}  \bibinfo{volume}{34} (\bibinfo{year}{2021}), \bibinfo{pages}{26924--26936}.
\newblock


\bibitem[Dai et~al\mbox{.}(2023)]%
        {chatgptrs}
\bibfield{author}{\bibinfo{person}{Sunhao Dai}, \bibinfo{person}{Ninglu Shao}, \bibinfo{person}{Haiyuan Zhao}, \bibinfo{person}{Weijie Yu}, \bibinfo{person}{Zihua Si}, \bibinfo{person}{Chen Xu}, \bibinfo{person}{Zhongxiang Sun}, \bibinfo{person}{Xiao Zhang}, {and} \bibinfo{person}{Jun Xu}.} \bibinfo{year}{2023}\natexlab{}.
\newblock \showarticletitle{Uncovering ChatGPT’s Capabilities in Recommender Systems}. In \bibinfo{booktitle}{\emph{Proceedings of the 17th ACM Conference on Recommender Systems (RecSys)}}. \bibinfo{pages}{1126–1132}.
\newblock


\bibitem[Dai et~al\mbox{.}(2024)]%
        {icdemusic}
\bibfield{author}{\bibinfo{person}{Sunhao Dai}, \bibinfo{person}{Ninglu Shao}, \bibinfo{person}{Jieming Zhu}, \bibinfo{person}{Xiao Zhang}, \bibinfo{person}{Zhenhua Dong}, \bibinfo{person}{Jun Xu}, \bibinfo{person}{Quanyu Dai}, {and} \bibinfo{person}{Ji-Rong Wen}.} \bibinfo{year}{2024}\natexlab{}.
\newblock \showarticletitle{Modeling User Attention in Music Recommendation}. In \bibinfo{booktitle}{\emph{2024 IEEE 40th International Conference on Data Engineering (ICDE)}}. \bibinfo{pages}{761--774}.
\newblock


\bibitem[Du et~al\mbox{.}(2021)]%
        {du2021modeling}
\bibfield{author}{\bibinfo{person}{Yingpeng Du}, \bibinfo{person}{Hongzhi Liu}, {and} \bibinfo{person}{Zhonghai Wu}.} \bibinfo{year}{2021}\natexlab{}.
\newblock \showarticletitle{Modeling multi-factor and multi-faceted preferences over sequential networks for next item recommendation}. In \bibinfo{booktitle}{\emph{Machine Learning and Knowledge Discovery in Databases. Research Track: European Conference, ECML PKDD 2021, Spain, 2021, Proceedings, Part II}}. \bibinfo{pages}{516--531}.
\newblock


\bibitem[Du et~al\mbox{.}(2024a)]%
        {du2023enhancing}
\bibfield{author}{\bibinfo{person}{Yingpeng Du}, \bibinfo{person}{Di Luo}, \bibinfo{person}{Rui Yan}, \bibinfo{person}{Xiaopei Wang}, \bibinfo{person}{Hongzhi Liu}, \bibinfo{person}{Hengshu Zhu}, \bibinfo{person}{Yang Song}, {and} \bibinfo{person}{Jie Zhang}.} \bibinfo{year}{2024}\natexlab{a}.
\newblock \showarticletitle{Enhancing Job Recommendation through LLM-Based Generative Adversarial Networks}.
\newblock \bibinfo{journal}{\emph{Proceedings of the AAAI Conference on Artificial Intelligence (AAAI)}} (\bibinfo{year}{2024}), \bibinfo{pages}{8363--8371}.
\newblock


\bibitem[Du et~al\mbox{.}(2024b)]%
        {dismir}
\bibfield{author}{\bibinfo{person}{Yingpeng Du}, \bibinfo{person}{Ziyan Wang}, \bibinfo{person}{Zhu Sun}, \bibinfo{person}{Yining Ma}, \bibinfo{person}{Hongzhi Liu}, {and} \bibinfo{person}{Jie Zhang}.} \bibinfo{year}{2024}\natexlab{b}.
\newblock \showarticletitle{Disentangled Multi-interest Representation Learning for Sequential Recommendation}. In \bibinfo{booktitle}{\emph{Proceedings of the 30th ACM SIGKDD Conference on Knowledge Discovery and Data Mining (KDD)}}. \bibinfo{pages}{677–688}.
\newblock


\bibitem[Geng et~al\mbox{.}(2022a)]%
        {p5}
\bibfield{author}{\bibinfo{person}{Shijie Geng}, \bibinfo{person}{Shuchang Liu}, \bibinfo{person}{Zuohui Fu}, \bibinfo{person}{Yingqiang Ge}, {and} \bibinfo{person}{Yongfeng Zhang}.} \bibinfo{year}{2022}\natexlab{a}.
\newblock \showarticletitle{Recommendation as Language Processing (RLP): A Unified Pretrain, Personalized Prompt \& Predict Paradigm (P5)}. In \bibinfo{booktitle}{\emph{Proceedings of the 16th ACM Conference on Recommender Systems (RecSys)}}. \bibinfo{pages}{299–315}.
\newblock


\bibitem[Geng et~al\mbox{.}(2022b)]%
        {geng2022recommendation}
\bibfield{author}{\bibinfo{person}{Shijie Geng}, \bibinfo{person}{Shuchang Liu}, \bibinfo{person}{Zuohui Fu}, \bibinfo{person}{Yingqiang Ge}, {and} \bibinfo{person}{Yongfeng Zhang}.} \bibinfo{year}{2022}\natexlab{b}.
\newblock \showarticletitle{Recommendation as Language Processing (RLP): A Unified Pretrain, Personalized Prompt \& Predict Paradigm (p5)}. In \bibinfo{booktitle}{\emph{Proceedings of the 16th ACM Conference on Recommender Systems (RecSys)}}. \bibinfo{pages}{299--315}.
\newblock


\bibitem[Hamilton et~al\mbox{.}(2017)]%
        {hamilton2017inductive}
\bibfield{author}{\bibinfo{person}{Will Hamilton}, \bibinfo{person}{Zhitao Ying}, {and} \bibinfo{person}{Jure Leskovec}.} \bibinfo{year}{2017}\natexlab{}.
\newblock \showarticletitle{Inductive representation learning on large graphs}.
\newblock \bibinfo{journal}{\emph{Advances in Neural Information Processing Systems (NeurIPS)}}  \bibinfo{volume}{30} (\bibinfo{year}{2017}).
\newblock


\bibitem[Harte et~al\mbox{.}(2023)]%
        {llmbert4rec}
\bibfield{author}{\bibinfo{person}{Jesse Harte}, \bibinfo{person}{Wouter Zorgdrager}, \bibinfo{person}{Panos Louridas}, \bibinfo{person}{Asterios Katsifodimos}, \bibinfo{person}{Dietmar Jannach}, {and} \bibinfo{person}{Marios Fragkoulis}.} \bibinfo{year}{2023}\natexlab{}.
\newblock \showarticletitle{Leveraging Large Language Models for Sequential Recommendation}. In \bibinfo{booktitle}{\emph{Proceedings of the 17th ACM Conference on Recommender Systems (RecSys)}}.
\newblock


\bibitem[He et~al\mbox{.}(2020)]%
        {lightgcn}
\bibfield{author}{\bibinfo{person}{Xiangnan He}, \bibinfo{person}{Kuan Deng}, \bibinfo{person}{Xiang Wang}, \bibinfo{person}{Yan Li}, \bibinfo{person}{Yongdong Zhang}, {and} \bibinfo{person}{Meng Wang}.} \bibinfo{year}{2020}\natexlab{}.
\newblock \showarticletitle{Lightgcn: Simplifying and powering graph convolution network for recommendation}. In \bibinfo{booktitle}{\emph{Proceedings of the 43rd International ACM SIGIR conference on research and development in Information Retrieval}}. \bibinfo{pages}{639--648}.
\newblock


\bibitem[He et~al\mbox{.}(2017)]%
        {ncf}
\bibfield{author}{\bibinfo{person}{Xiangnan He}, \bibinfo{person}{Lizi Liao}, \bibinfo{person}{Hanwang Zhang}, \bibinfo{person}{Liqiang Nie}, \bibinfo{person}{Xia Hu}, {and} \bibinfo{person}{Tat-Seng Chua}.} \bibinfo{year}{2017}\natexlab{}.
\newblock \showarticletitle{Neural Collaborative Filtering}. In \bibinfo{booktitle}{\emph{Proceedings of the 26th International Conference on World Wide Web (WWW)}}. \bibinfo{pages}{173–182}.
\newblock


\bibitem[Hidasi et~al\mbox{.}(2016)]%
        {gru4rec}
\bibfield{author}{\bibinfo{person}{Balazs Hidasi}, \bibinfo{person}{Alexandros Karatzoglou}, \bibinfo{person}{Linas Baltrunas}, {and} \bibinfo{person}{Domonkos Tikk}.} \bibinfo{year}{2016}\natexlab{}.
\newblock \showarticletitle{Session-based recommendations with recurrent neural networks}. In \bibinfo{booktitle}{\emph{International Conference on Learning Representations (ICLR)}}.
\newblock


\bibitem[Hou et~al\mbox{.}(2022)]%
        {unisrec}
\bibfield{author}{\bibinfo{person}{Yupeng Hou}, \bibinfo{person}{Shanlei Mu}, \bibinfo{person}{Wayne~Xin Zhao}, \bibinfo{person}{Yaliang Li}, \bibinfo{person}{Bolin Ding}, {and} \bibinfo{person}{Ji-Rong Wen}.} \bibinfo{year}{2022}\natexlab{}.
\newblock \showarticletitle{Towards Universal Sequence Representation Learning for Recommender Systems}. In \bibinfo{booktitle}{\emph{Proceedings of the 28th ACM SIGKDD Conference on Knowledge Discovery and Data Mining (KDD)}}. \bibinfo{pages}{585–593}.
\newblock


\bibitem[Hou et~al\mbox{.}(2024)]%
        {llmrank}
\bibfield{author}{\bibinfo{person}{Yupeng Hou}, \bibinfo{person}{Junjie Zhang}, \bibinfo{person}{Zihan Lin}, \bibinfo{person}{Hongyu Lu}, \bibinfo{person}{Ruobing Xie}, \bibinfo{person}{Julian McAuley}, {and} \bibinfo{person}{Wayne~Xin Zhao}.} \bibinfo{year}{2024}\natexlab{}.
\newblock \showarticletitle{Large Language Models are Zero-Shot Rankers for Recommender Systems}. In \bibinfo{booktitle}{\emph{European Conference on Information Retrieval (ECIR)}}. \bibinfo{pages}{364--381}.
\newblock


\bibitem[Hu et~al\mbox{.}(2022)]%
        {lora}
\bibfield{author}{\bibinfo{person}{Edward~J Hu}, \bibinfo{person}{Yelong Shen}, \bibinfo{person}{Phillip Wallis}, \bibinfo{person}{Zeyuan Allen-Zhu}, \bibinfo{person}{Yuanzhi Li}, \bibinfo{person}{Shean Wang}, \bibinfo{person}{Lu Wang}, {and} \bibinfo{person}{Weizhu Chen}.} \bibinfo{year}{2022}\natexlab{}.
\newblock \showarticletitle{LoRA: Low-Rank Adaptation of Large Language Models}. In \bibinfo{booktitle}{\emph{International Conference on Learning Representations (ICLR)}}.
\newblock


\bibitem[Jaeri et~al\mbox{.}(2024)]%
        {bam}
\bibfield{author}{\bibinfo{person}{Lee Jaeri}, \bibinfo{person}{Yun Jeongin}, {and} \bibinfo{person}{Kang U}.} \bibinfo{year}{2024}\natexlab{}.
\newblock \showarticletitle{BAM: Enhancing Recommendation with Backward-Aware Mechanisms}. In \bibinfo{booktitle}{\emph{Proceedings of the 30th ACM SIGKDD Conference on Knowledge Discovery and Data Mining (KDD)}}.
\newblock


\bibitem[Ji et~al\mbox{.}(2024)]%
        {genrec}
\bibfield{author}{\bibinfo{person}{Jianchao Ji}, \bibinfo{person}{Zelong Li}, \bibinfo{person}{Shuyuan Xu}, \bibinfo{person}{Wenyue Hua}, \bibinfo{person}{Yingqiang Ge}, \bibinfo{person}{Juntao Tan}, {and} \bibinfo{person}{Yongfeng Zhang}.} \bibinfo{year}{2024}\natexlab{}.
\newblock \showarticletitle{GenRec: Large Language Model for Generative Recommendation}. In \bibinfo{booktitle}{\emph{European Conference on Information Retrieval (ECIR)}}. \bibinfo{pages}{494--502}.
\newblock


\bibitem[Kang and McAuley(2018)]%
        {sasrec}
\bibfield{author}{\bibinfo{person}{Wang-Cheng Kang} {and} \bibinfo{person}{Julian McAuley}.} \bibinfo{year}{2018}\natexlab{}.
\newblock \showarticletitle{Self-Attentive Sequential Recommendation}. In \bibinfo{booktitle}{\emph{2018 IEEE International Conference on Data Mining (ICDM)}}. \bibinfo{pages}{197--206}.
\newblock


\bibitem[Khuller et~al\mbox{.}(1999)]%
        {mcp}
\bibfield{author}{\bibinfo{person}{Samir Khuller}, \bibinfo{person}{Anna Moss}, {and} \bibinfo{person}{Joseph~Seffi Naor}.} \bibinfo{year}{1999}\natexlab{}.
\newblock \showarticletitle{The budgeted maximum coverage problem}.
\newblock \bibinfo{journal}{\emph{Information processing letters}} \bibinfo{volume}{70}, \bibinfo{number}{1} (\bibinfo{year}{1999}), \bibinfo{pages}{39--45}.
\newblock


\bibitem[Li et~al\mbox{.}(2019)]%
        {mind}
\bibfield{author}{\bibinfo{person}{Chao Li}, \bibinfo{person}{Zhiyuan Liu}, \bibinfo{person}{Mengmeng Wu}, \bibinfo{person}{Yuchi Xu}, \bibinfo{person}{Huan Zhao}, \bibinfo{person}{Pipei Huang}, \bibinfo{person}{Guoliang Kang}, \bibinfo{person}{Qiwei Chen}, \bibinfo{person}{Wei Li}, {and} \bibinfo{person}{Dik~Lun Lee}.} \bibinfo{year}{2019}\natexlab{}.
\newblock \showarticletitle{Multi-Interest Network with Dynamic Routing for Recommendation at Tmall}. In \bibinfo{booktitle}{\emph{Proceedings of the 28th ACM International Conference on Information and Knowledge Management (CIKM)}}. \bibinfo{pages}{2615–2623}.
\newblock


\bibitem[Li et~al\mbox{.}(2017)]%
        {narm}
\bibfield{author}{\bibinfo{person}{Jing Li}, \bibinfo{person}{Pengjie Ren}, \bibinfo{person}{Zhumin Chen}, \bibinfo{person}{Zhaochun Ren}, \bibinfo{person}{Tao Lian}, {and} \bibinfo{person}{Jun Ma}.} \bibinfo{year}{2017}\natexlab{}.
\newblock \showarticletitle{Neural Attentive Session-based Recommendation}. In \bibinfo{booktitle}{\emph{Proceedings of the 26th ACM International Conference on Information and Knowledge Management (CIKM)}}. \bibinfo{pages}{1419–1428}.
\newblock


\bibitem[Lin et~al\mbox{.}(2024)]%
        {transrec}
\bibfield{author}{\bibinfo{person}{Xinyu Lin}, \bibinfo{person}{Wenjie Wang}, \bibinfo{person}{Yongqi Li}, \bibinfo{person}{Fuli Feng}, \bibinfo{person}{See-Kiong Ng}, {and} \bibinfo{person}{Tat-Seng Chua}.} \bibinfo{year}{2024}\natexlab{}.
\newblock \showarticletitle{Bridging Items and Language: A Transition Paradigm for Large Language Model-Based Recommendation}. In \bibinfo{booktitle}{\emph{Proceedings of the 30th ACM SIGKDD Conference on Knowledge Discovery and Data Mining (KDD)}}. \bibinfo{pages}{1816–1826}.
\newblock


\bibitem[Liu et~al\mbox{.}(2022)]%
        {kgmulti}
\bibfield{author}{\bibinfo{person}{Danyang Liu}, \bibinfo{person}{Yuji Yang}, \bibinfo{person}{Mengdi Zhang}, \bibinfo{person}{Wei Wu}, \bibinfo{person}{Xing Xie}, {and} \bibinfo{person}{Guangzhong Sun}.} \bibinfo{year}{2022}\natexlab{}.
\newblock \showarticletitle{Knowledge Enhanced Multi-Interest Network for the Generation of Recommendation Candidates}. In \bibinfo{booktitle}{\emph{Proceedings of the 31st ACM International Conference on Information \& Knowledge Management (CIKM)}}. \bibinfo{pages}{3322–3331}.
\newblock


\bibitem[Liu et~al\mbox{.}(2024a)]%
        {once}
\bibfield{author}{\bibinfo{person}{Qijiong Liu}, \bibinfo{person}{Nuo Chen}, \bibinfo{person}{Tetsuya Sakai}, {and} \bibinfo{person}{Xiao-Ming Wu}.} \bibinfo{year}{2024}\natexlab{a}.
\newblock \showarticletitle{ONCE: Boosting Content-based Recommendation with Both Open- and Closed-source Large Language Models}. In \bibinfo{booktitle}{\emph{Proceedings of the 17th ACM International Conference on Web Search and Data Mining (WSDM)}}. \bibinfo{pages}{452–461}.
\newblock


\bibitem[Liu et~al\mbox{.}(2024b)]%
        {liu2024once}
\bibfield{author}{\bibinfo{person}{Qijiong Liu}, \bibinfo{person}{Nuo Chen}, \bibinfo{person}{Tetsuya Sakai}, {and} \bibinfo{person}{Xiao-Ming Wu}.} \bibinfo{year}{2024}\natexlab{b}.
\newblock \showarticletitle{Once: Boosting Content-based Recommendation with Both Open-and Closed-source Large Language Models}. In \bibinfo{booktitle}{\emph{Proceedings of the 17th ACM International Conference on Web Search and Data Mining (WSDM)}}. \bibinfo{pages}{452--461}.
\newblock


\bibitem[Liu et~al\mbox{.}(2024c)]%
        {multi-attr}
\bibfield{author}{\bibinfo{person}{Yaokun Liu}, \bibinfo{person}{Xiaowang Zhang}, \bibinfo{person}{Minghui Zou}, {and} \bibinfo{person}{Zhiyong Feng}.} \bibinfo{year}{2024}\natexlab{c}.
\newblock \showarticletitle{Attribute Simulation for Item Embedding Enhancement in Multi-interest Recommendation}. In \bibinfo{booktitle}{\emph{Proceedings of the 17th ACM International Conference on Web Search and Data Mining (WSDM)}}. \bibinfo{pages}{482–491}.
\newblock


\bibitem[Ma et~al\mbox{.}(2020)]%
        {ssl}
\bibfield{author}{\bibinfo{person}{Jianxin Ma}, \bibinfo{person}{Chang Zhou}, \bibinfo{person}{Hongxia Yang}, \bibinfo{person}{Peng Cui}, \bibinfo{person}{Xin Wang}, {and} \bibinfo{person}{Wenwu Zhu}.} \bibinfo{year}{2020}\natexlab{}.
\newblock \showarticletitle{Disentangled Self-Supervision in Sequential Recommenders}. In \bibinfo{booktitle}{\emph{Proceedings of the 26th ACM SIGKDD International Conference on Knowledge Discovery \& Data Mining (KDD)}}. \bibinfo{pages}{483–491}.
\newblock


\bibitem[Meng et~al\mbox{.}(2023)]%
        {meng2023coarse}
\bibfield{author}{\bibinfo{person}{Chang Meng}, \bibinfo{person}{Ziqi Zhao}, \bibinfo{person}{Wei Guo}, \bibinfo{person}{Yingxue Zhang}, \bibinfo{person}{Haolun Wu}, \bibinfo{person}{Chen Gao}, \bibinfo{person}{Dong Li}, \bibinfo{person}{Xiu Li}, {and} \bibinfo{person}{Ruiming Tang}.} \bibinfo{year}{2023}\natexlab{}.
\newblock \showarticletitle{Coarse-to-fine knowledge-enhanced multi-interest learning framework for multi-behavior recommendation}.
\newblock \bibinfo{journal}{\emph{ACM Transactions on Information Systems (TIS)}} \bibinfo{volume}{42}, \bibinfo{number}{1} (\bibinfo{year}{2023}), \bibinfo{pages}{1--27}.
\newblock


\bibitem[Ni et~al\mbox{.}(2019)]%
        {amazonreview}
\bibfield{author}{\bibinfo{person}{Jianmo Ni}, \bibinfo{person}{Jiacheng Li}, {and} \bibinfo{person}{Julian McAuley}.} \bibinfo{year}{2019}\natexlab{}.
\newblock \showarticletitle{Justifying Recommendations using Distantly-Labeled Reviews and Fine-Grained Aspects}. In \bibinfo{booktitle}{\emph{Proceedings of the 2019 Conference on Empirical Methods in Natural Language Processing and the 9th International Joint Conference on Natural Language Processing (EMNLP-IJCNLP)}}, \bibfield{editor}{\bibinfo{person}{Kentaro Inui}, \bibinfo{person}{Jing Jiang}, \bibinfo{person}{Vincent Ng}, {and} \bibinfo{person}{Xiaojun Wan}} (Eds.). \bibinfo{pages}{188--197}.
\newblock


\bibitem[Qiao et~al\mbox{.}(2024)]%
        {qiao2024llm}
\bibfield{author}{\bibinfo{person}{Shutong Qiao}, \bibinfo{person}{Chen Gao}, \bibinfo{person}{Yong Li}, {and} \bibinfo{person}{Hongzhi Yin}.} \bibinfo{year}{2024}\natexlab{}.
\newblock \showarticletitle{LLM-assisted Explicit and Implicit Multi-interest Learning Framework for Sequential Recommendation}.
\newblock \bibinfo{journal}{\emph{arXiv preprint arXiv:2411.09410}} (\bibinfo{year}{2024}).
\newblock


\bibitem[Qiu et~al\mbox{.}(2021)]%
        {ubert}
\bibfield{author}{\bibinfo{person}{Zhaopeng Qiu}, \bibinfo{person}{Xian Wu}, \bibinfo{person}{Jingyue Gao}, {and} \bibinfo{person}{Wei Fan}.} \bibinfo{year}{2021}\natexlab{}.
\newblock \showarticletitle{U-BERT: Pre-training user representations for improved recommendation}. In \bibinfo{booktitle}{\emph{Proceedings of the AAAI Conference on Artificial Intelligence (AAAI)}}, Vol.~\bibinfo{volume}{35}. \bibinfo{pages}{4320--4327}.
\newblock


\bibitem[Qu et~al\mbox{.}(2024)]%
        {elephant}
\bibfield{author}{\bibinfo{person}{Zekai Qu}, \bibinfo{person}{Ruobing Xie}, \bibinfo{person}{Chaojun Xiao}, \bibinfo{person}{Zhanhui Kang}, {and} \bibinfo{person}{Xingwu Sun}.} \bibinfo{year}{2024}\natexlab{}.
\newblock \showarticletitle{The Elephant in the Room: Rethinking the Usage of Pre-trained Language Model in Sequential Recommendation}. In \bibinfo{booktitle}{\emph{Proceedings of the 18th ACM Conference on Recommender Systems (RecSys)}}. \bibinfo{pages}{53–62}.
\newblock


\bibitem[Ren et~al\mbox{.}(2024)]%
        {ren2024representation}
\bibfield{author}{\bibinfo{person}{Xubin Ren}, \bibinfo{person}{Wei Wei}, \bibinfo{person}{Lianghao Xia}, \bibinfo{person}{Lixin Su}, \bibinfo{person}{Suqi Cheng}, \bibinfo{person}{Junfeng Wang}, \bibinfo{person}{Dawei Yin}, {and} \bibinfo{person}{Chao Huang}.} \bibinfo{year}{2024}\natexlab{}.
\newblock \showarticletitle{RLMRec: Representation Learning with Large Language Models for Recommendation}. In \bibinfo{booktitle}{\emph{Proceedings of the ACM on Web Conference 2024 (TheWebConf)}}. \bibinfo{pages}{3464--3475}.
\newblock


\bibitem[Sabour et~al\mbox{.}(2017)]%
        {capsulenet}
\bibfield{author}{\bibinfo{person}{Sara Sabour}, \bibinfo{person}{Nicholas Frosst}, {and} \bibinfo{person}{Geoffrey~E. Hinton}.} \bibinfo{year}{2017}\natexlab{}.
\newblock \showarticletitle{Dynamic routing between capsules}. In \bibinfo{booktitle}{\emph{Proceedings of the 31st International Conference on Neural Information Processing Systems (NeurIPS)}}. \bibinfo{pages}{3859–3869}.
\newblock


\bibitem[Sanner et~al\mbox{.}(2023)]%
        {icl-llmrec}
\bibfield{author}{\bibinfo{person}{Scott Sanner}, \bibinfo{person}{Krisztian Balog}, \bibinfo{person}{Filip Radlinski}, \bibinfo{person}{Ben Wedin}, {and} \bibinfo{person}{Lucas Dixon}.} \bibinfo{year}{2023}\natexlab{}.
\newblock \showarticletitle{Large Language Models are Competitive Near Cold-start Recommenders for Language- and Item-based Preferences}. In \bibinfo{booktitle}{\emph{Proceedings of the 17th ACM Conference on Recommender Systems (RecSys)}}. \bibinfo{pages}{890–896}.
\newblock


\bibitem[Sun et~al\mbox{.}(2019)]%
        {bert4rec}
\bibfield{author}{\bibinfo{person}{Fei Sun}, \bibinfo{person}{Jun Liu}, \bibinfo{person}{Jian Wu}, \bibinfo{person}{Changhua Pei}, \bibinfo{person}{Xiao Lin}, \bibinfo{person}{Wenwu Ou}, {and} \bibinfo{person}{Peng Jiang}.} \bibinfo{year}{2019}\natexlab{}.
\newblock \showarticletitle{BERT4Rec: Sequential Recommendation with Bidirectional Encoder Representations from Transformer}. In \bibinfo{booktitle}{\emph{Proceedings of the 28th ACM International Conference on Information and Knowledge Management (CIKM)}}. \bibinfo{pages}{1441–1450}.
\newblock


\bibitem[Sun et~al\mbox{.}(2023)]%
        {social}
\bibfield{author}{\bibinfo{person}{Youchen Sun}, \bibinfo{person}{Zhu Sun}, \bibinfo{person}{Xiao Sha}, \bibinfo{person}{Jie Zhang}, {and} \bibinfo{person}{Yew~Soon Ong}.} \bibinfo{year}{2023}\natexlab{}.
\newblock \showarticletitle{Disentangling Motives behind Item Consumption and Social Connection for Mutually-enhanced Joint Prediction}. In \bibinfo{booktitle}{\emph{Proceedings of the 17th ACM Conference on Recommender Systems (RecSys)}}. \bibinfo{pages}{613–624}.
\newblock


\bibitem[Tian et~al\mbox{.}(2022)]%
        {multimeet1}
\bibfield{author}{\bibinfo{person}{Yu Tian}, \bibinfo{person}{Jianxin Chang}, \bibinfo{person}{Yanan Niu}, \bibinfo{person}{Yang Song}, {and} \bibinfo{person}{Chenliang Li}.} \bibinfo{year}{2022}\natexlab{}.
\newblock \showarticletitle{When Multi-Level Meets Multi-Interest: A Multi-Grained Neural Model for Sequential Recommendation}. In \bibinfo{booktitle}{\emph{Proceedings of the 45th International ACM SIGIR Conference on Research and Development in Information Retrieval (SIGIR)}}. \bibinfo{pages}{1632–1641}.
\newblock


\bibitem[Wang et~al\mbox{.}(2022b)]%
        {narco}
\bibfield{author}{\bibinfo{person}{Runzhong Wang}, \bibinfo{person}{Li Shen}, \bibinfo{person}{Yiting Chen}, \bibinfo{person}{Xiaokang Yang}, \bibinfo{person}{Dacheng Tao}, {and} \bibinfo{person}{Junchi Yan}.} \bibinfo{year}{2022}\natexlab{b}.
\newblock \showarticletitle{Towards one-shot neural combinatorial solvers: Theoretical and empirical notes on the cardinality-constrained case}. In \bibinfo{booktitle}{\emph{The 11th International Conference on Learning Representations (ICLR)}}.
\newblock


\bibitem[Wang et~al\mbox{.}(2022a)]%
        {wang2022disentangled}
\bibfield{author}{\bibinfo{person}{Xin Wang}, \bibinfo{person}{Hong Chen}, \bibinfo{person}{Yuwei Zhou}, \bibinfo{person}{Jianxin Ma}, {and} \bibinfo{person}{Wenwu Zhu}.} \bibinfo{year}{2022}\natexlab{a}.
\newblock \showarticletitle{Disentangled representation learning for recommendation}.
\newblock \bibinfo{journal}{\emph{IEEE Transactions on Pattern Analysis and Machine Intelligence (TPAMI)}} \bibinfo{volume}{45}, \bibinfo{number}{1} (\bibinfo{year}{2022}), \bibinfo{pages}{408--424}.
\newblock


\bibitem[Wang et~al\mbox{.}(2021)]%
        {wang2021multimodal}
\bibfield{author}{\bibinfo{person}{Xin Wang}, \bibinfo{person}{Hong Chen}, {and} \bibinfo{person}{Wenwu Zhu}.} \bibinfo{year}{2021}\natexlab{}.
\newblock \showarticletitle{Multimodal disentangled representation for recommendation}. In \bibinfo{booktitle}{\emph{2021 IEEE International Conference on Multimedia and Expo (ICME)}}. IEEE, \bibinfo{pages}{1--6}.
\newblock


\bibitem[Wang et~al\mbox{.}(2024)]%
        {recmind}
\bibfield{author}{\bibinfo{person}{Yancheng Wang}, \bibinfo{person}{Ziyan Jiang}, \bibinfo{person}{Zheng Chen}, \bibinfo{person}{Fan Yang}, \bibinfo{person}{Yingxue Zhou}, \bibinfo{person}{Eunah Cho}, \bibinfo{person}{Xing Fan}, \bibinfo{person}{Yanbin Lu}, \bibinfo{person}{Xiaojiang Huang}, {and} \bibinfo{person}{Yingzhen Yang}.} \bibinfo{year}{2024}\natexlab{}.
\newblock \showarticletitle{{R}ec{M}ind: Large Language Model Powered Agent For Recommendation}. In \bibinfo{booktitle}{\emph{Findings of the Association for Computational Linguistics: NAACL 2024}}.
\newblock


\bibitem[Wang et~al\mbox{.}(2023)]%
        {wang2023intent}
\bibfield{author}{\bibinfo{person}{Yuling Wang}, \bibinfo{person}{Xiao Wang}, \bibinfo{person}{Xiangzhou Huang}, \bibinfo{person}{Yanhua Yu}, \bibinfo{person}{Haoyang Li}, \bibinfo{person}{Mengdi Zhang}, \bibinfo{person}{Zirui Guo}, {and} \bibinfo{person}{Wei Wu}.} \bibinfo{year}{2023}\natexlab{}.
\newblock \showarticletitle{Intent-aware recommendation via disentangled graph contrastive learning}. In \bibinfo{booktitle}{\emph{Proceedings of the 32nd International Joint Conference on Artificial Intelligence (IJCAI)}}. \bibinfo{pages}{2343--2351}.
\newblock


\bibitem[Wei et~al\mbox{.}(2024)]%
        {wei2024llmrec}
\bibfield{author}{\bibinfo{person}{Wei Wei}, \bibinfo{person}{Xubin Ren}, \bibinfo{person}{Jiabin Tang}, \bibinfo{person}{Qinyong Wang}, \bibinfo{person}{Lixin Su}, \bibinfo{person}{Suqi Cheng}, \bibinfo{person}{Junfeng Wang}, \bibinfo{person}{Dawei Yin}, {and} \bibinfo{person}{Chao Huang}.} \bibinfo{year}{2024}\natexlab{}.
\newblock \showarticletitle{LLMRec: Large Language Models with Graph Augmentation for Recommendation}. In \bibinfo{booktitle}{\emph{Proceedings of the 17th ACM International Conference on Web Search and Data Mining (WSDM)}}. \bibinfo{pages}{806--815}.
\newblock


\bibitem[Wu et~al\mbox{.}(2024)]%
        {llmrecsurvey}
\bibfield{author}{\bibinfo{person}{Likang Wu}, \bibinfo{person}{Zhi Zheng}, \bibinfo{person}{Zhaopeng Qiu}, \bibinfo{person}{Hao Wang}, \bibinfo{person}{Hongchao Gu}, \bibinfo{person}{Tingjia Shen}, \bibinfo{person}{Chuan Qin}, \bibinfo{person}{Chen Zhu}, \bibinfo{person}{Hengshu Zhu}, \bibinfo{person}{Qi Liu}, \bibinfo{person}{Hui Xiong}, {and} \bibinfo{person}{Enhong Chen}.} \bibinfo{year}{2024}\natexlab{}.
\newblock \showarticletitle{A survey on large language models for recommendation}.
\newblock \bibinfo{journal}{\emph{World Wide Web}}  \bibinfo{volume}{27} (\bibinfo{year}{2024}), \bibinfo{pages}{60}.
\newblock


\bibitem[Xiao et~al\mbox{.}(2020)]%
        {attenmulti}
\bibfield{author}{\bibinfo{person}{Zhibo Xiao}, \bibinfo{person}{Luwei Yang}, \bibinfo{person}{Wen Jiang}, \bibinfo{person}{Yi Wei}, \bibinfo{person}{Yi Hu}, {and} \bibinfo{person}{Hao Wang}.} \bibinfo{year}{2020}\natexlab{}.
\newblock \showarticletitle{Deep Multi-Interest Network for Click-through Rate Prediction}. In \bibinfo{booktitle}{\emph{Proceedings of the 29th ACM International Conference on Information \& Knowledge Management (CIKM)}}. \bibinfo{pages}{2265–2268}.
\newblock


\bibitem[Xie et~al\mbox{.}(2022)]%
        {aug}
\bibfield{author}{\bibinfo{person}{Xu Xie}, \bibinfo{person}{Fei Sun}, \bibinfo{person}{Zhaoyang Liu}, \bibinfo{person}{Shiwen Wu}, \bibinfo{person}{Jinyang Gao}, \bibinfo{person}{Jiandong Zhang}, \bibinfo{person}{Bolin Ding}, {and} \bibinfo{person}{Bin Cui}.} \bibinfo{year}{2022}\natexlab{}.
\newblock \showarticletitle{Contrastive Learning for Sequential Recommendation}. In \bibinfo{booktitle}{\emph{IEEE 38th International Conference on Data Engineering (ICDE)}}. \bibinfo{pages}{1259--1273}.
\newblock


\bibitem[Xie et~al\mbox{.}(2023)]%
        {remi}
\bibfield{author}{\bibinfo{person}{Yueqi Xie}, \bibinfo{person}{Jingqi Gao}, \bibinfo{person}{Peilin Zhou}, \bibinfo{person}{Qichen Ye}, \bibinfo{person}{Yining Hua}, \bibinfo{person}{Jae~Boum Kim}, \bibinfo{person}{Fangzhao Wu}, {and} \bibinfo{person}{Sunghun Kim}.} \bibinfo{year}{2023}\natexlab{}.
\newblock \showarticletitle{Rethinking Multi-Interest Learning for Candidate Matching in Recommender Systems}. In \bibinfo{booktitle}{\emph{Proceedings of the 17th ACM Conference on Recommender Systems (RecSys)}}. \bibinfo{pages}{283–293}.
\newblock


\bibitem[Zhang et~al\mbox{.}(2024)]%
        {icdegraph}
\bibfield{author}{\bibinfo{person}{Qianru Zhang}, \bibinfo{person}{Lianghao Xia}, \bibinfo{person}{Xuheng Cai}, \bibinfo{person}{Siu-Ming Yiu}, \bibinfo{person}{Chao Huang}, {and} \bibinfo{person}{Christian~S. Jensen}.} \bibinfo{year}{2024}\natexlab{}.
\newblock \showarticletitle{Graph Augmentation for Recommendation}. In \bibinfo{booktitle}{\emph{IEEE 40th International Conference on Data Engineering (ICDE)}}. \bibinfo{pages}{557--569}.
\newblock


\bibitem[Zhang et~al\mbox{.}(2022)]%
        {re4}
\bibfield{author}{\bibinfo{person}{Shengyu Zhang}, \bibinfo{person}{Lingxiao Yang}, \bibinfo{person}{Dong Yao}, \bibinfo{person}{Yujie Lu}, \bibinfo{person}{Fuli Feng}, \bibinfo{person}{Zhou Zhao}, \bibinfo{person}{Tat{-}Seng Chua}, {and} \bibinfo{person}{Fei Wu}.} \bibinfo{year}{2022}\natexlab{}.
\newblock \showarticletitle{Re4: Learning to Re-contrast, Re-attend, Re-construct for Multi-interest Recommendation}. In \bibinfo{booktitle}{\emph{The {ACM} Web Conference (TheWebConf)}}. \bibinfo{pages}{2216--2226}.
\newblock


\bibitem[Zhao et~al\mbox{.}(2021)]%
        {varsa}
\bibfield{author}{\bibinfo{person}{Jing Zhao}, \bibinfo{person}{Pengpeng Zhao}, \bibinfo{person}{Lei Zhao}, \bibinfo{person}{Yanchi Liu}, \bibinfo{person}{Victor~S. Sheng}, {and} \bibinfo{person}{Xiaofang Zhou}.} \bibinfo{year}{2021}\natexlab{}.
\newblock \showarticletitle{Variational Self-attention Network for Sequential Recommendation}. In \bibinfo{booktitle}{\emph{IEEE 37th International Conference on Data Engineering (ICDE)}}. \bibinfo{pages}{1559--1570}.
\newblock


\bibitem[Zheng et~al\mbox{.}(2023)]%
        {zheng2023generative}
\bibfield{author}{\bibinfo{person}{Zhi Zheng}, \bibinfo{person}{Zhaopeng Qiu}, \bibinfo{person}{Xiao Hu}, \bibinfo{person}{Likang Wu}, \bibinfo{person}{Hengshu Zhu}, {and} \bibinfo{person}{Hui Xiong}.} \bibinfo{year}{2023}\natexlab{}.
\newblock \showarticletitle{Generative Job Recommendations with Large Language Model}.
\newblock \bibinfo{journal}{\emph{arXiv preprint arXiv:2307.02157}} (\bibinfo{year}{2023}).
\newblock


\end{thebibliography}
\end{document}